\documentclass[traditabstract]{aa}

\usepackage{graphicx}
\usepackage{natbib}

\usepackage{piero-macro-private}

\begin{document}

\title{The 2-10 keV unabsorbed luminosity function of AGN\\ from the
LSS, CDFS, and COSMOS surveys
\thanks{Based on observations obtained with XMM-Newton, an ESA science mission with instruments and contributions directly funded by ESA member states and NASA.}
\thanks{The tables with the samples of the posterior probability distributions are only available in electronic form
at the CDS via anonymous ftp to cdsarc.u-strasbg.fr (130.79.128.5)
or via http://cdsweb.u-strasbg.fr/cgi-bin/qcat?J/A+A/
}
}

   \author{P.~Ranalli
          \inst{1,2,3}
          \and
          E.~Koulouridis\inst{1,4}
          \and
          I.~Georgantopoulos\inst{1}
          \and
          S.~Fotopoulou\inst{5}
          \and
          L.-T.~Hsu\inst{6}
          \and
          M.~Salvato\inst{6}
          \and
          A.~Comastri\inst{2}
          \and
          M.~Pierre\inst{4}
          \and
          N.~Cappelluti\inst{2,7}
          \and
          F.~J.~Carrera\inst{8}
          \and
          L.~Chiappetti\inst{9}
          \and
          N.~Clerc\inst{5}
          \and
          R.~Gilli\inst{2}
          \and
          K.~Iwasawa\inst{10}
          \and
          F.~Pacaud\inst{11}
          \and
          S.~Paltani\inst{5}
          \and
          E.~Plionis\inst{12}
          \and
          C.~Vignali\inst{13,2}
          }

   \institute{
     Institute for Astronomy, Astrophysics, Space Applications and
     Remote Sensing (IAASARS), National Observatory of
     Athens,  15236 Penteli, Greece
    \and
     INAF -- Osservatorio Astronomico di Bologna,
     via Ranzani 1, 40127 Bologna, Italy
    \and
    Lund Observatory,
    Department of Astronomy and Theoretical Physics, Lund
    University, Box 43, 22100 Lund, Sweden;
    \email{piero@astro.lu.se} 
     \and
     Service d'astrophysique, IRFU, CEA Saclay, France
     \and
     Geneva Observatory, University of Geneva, ch. des Maillettes 51, 1290, Versoix, Switzerland
     \and
     Max-Planck-Institut f\"ur extraterrestrische Physick, 85478
     Garching, Germany
     \and
     Department of Physics, Yale University, P.O. Box 208121, New Haven, CT 06520, USA
     \and
     Instituto de F\'{\i}sica de Cantabria (CSIC-UC), 39005 Santander, Spain
     \and
     INAF -- IASF Milano, via Bassini 15, 20133, Milano, Italy
     \and
     ICREA and Institut de Ci\`encies del Cosmos (ICC), Universitat de
     Barcelona, (IEEC-UB), Mart\'\i\ y Franqu\`es 1, 08028 Barcelona,
     Spain
     \and
     Argelander Institute for Astronomy, Bonn University, D-53121
     Bonn, Germany
     \and
     Physics Department, Aristotle University of Thessaloniki, 54124 Thessaloniki, Greece
     \and
     Universit\`a di Bologna, Dipartimento di Fisica e Astronomia,
     via Berti Pichat 6/2, 40127 Bologna, Italy
}

   \date{Received 2015-07-20; accepted 2015-12-17}

   \abstract{ The XMM-Large scale structure (XMM-LSS),
     XMM-Cosmological evolution survey (XMM-COSMOS), and XMM-\chandra\
     deep field south (XMM-CDFS) surveys are complementary in terms of sky
     coverage and depth. Together, they form a clean sample with the
     least possible variance in instrument effective areas and point
     spread function. Therefore this is one of the best samples
     available to determine the 2--10 keV luminosity function of
     active galactic nuclei (AGN) and their evolution.  The samples
     and the relevant corrections for incompleteness are described. A
     total of 2887 AGN is used to build the LF in the luminosity
     interval $10^{42}$--$10^{46}$ \ergs and in the redshift interval
     0.001-4.  A new method to correct for absorption by considering
     the probability distribution for the column density conditioned
     on the hardness ratio is presented.  The binned luminosity
     function and its evolution is determined with a variant of the
     Page-Carrera method, which is improved to include corrections for
     absorption and to account for the full probability distribution
     of photometric redshifts. Parametric models, namely a double
     power law with luminosity and density evolution (LADE) or
     luminosity-dependent density evolution (LDDE), are explored using
     Bayesian inference. We introduce the Watanabe-Akaike information
     criterion (WAIC) to compare the models and estimate their
     predictive power.  Our data are best described by the LADE model,
     as hinted by the WAIC indicator. We also explore the recently
     proposed 15-parameter extended LDDE model and find that this
     extension is not supported by our data.  The strength of our
     method is that it provides unabsorbed, non-parametric estimates,
     credible intervals for luminosity function parameters, and a
     model choice based on predictive power for future data.  }


     \authorrunning{P. Ranalli et al.}
     \titlerunning{The 2--10 keV unabsorbed luminosity function of AGN}

   \maketitle

\section{Introduction}
\label{sec:intro}

An accurate census of active galactic nuclei (AGN) is a central element in
understanding the cosmic history of accretion onto supermassive black
holes (BH). Black hole growth is in turn closely connected to star formation.
Scaling relations exist between the masses of the BH and of the bulge
of the host galaxies
\citep{magorrian1998,ferrares2000,gebhardt2000,marconi2003,haering2004,hopkins2007,kormendy2009,guetelkin2009,zubovas2012}.
On a larger scale, it has been recognised that BHs and their hosts have
been growing together for a large part of cosmic time
\citep{marconi2004,alexander2012} and that they exhibit a similar
downsizing trend, i.e.\ that more massive systems were formed
earlier than lower mass systems
\citep{cowie1996,ueda03,kodama2004,hasinger2005,fontanot2009,santini2009}.

Active galactic nuclei are the principal constituent of the extragalactic X-ray sky, and
their integrated contribution essentially builds up the X-ray cosmic
background \citep{sw89,comastri95}. Modelling the X-ray background
requires knowledge of the X-ray luminosity function (LF) of AGN,
their evolution, and the distribution of the column density of the
absorbing medium. The fraction of Compton-thick AGN with
column densities $N_H\gtrsim 10^{24}$ \cmq have especially notable
uncertainties \citep{gilli2007,treister2009,akylas2012}.

X-ray LFs started to be estimated as soon as AGN samples became
available \citep{maccacaro1983,maccacaro1984} and progressed with
\textit{Einstein} and ROSAT surveys
\citep{maccacaro1991,boyle1993,boyle1994,page1996}.  Among the recent
estimates in the 2--10 keV band, we mention
\citet{ueda03,lafranca2005,barger05,silverman2008,ebrero2009,yencho2009};
\citet[hereafter A10]{aird2010}; \citet[hereafter U14]{ueda2014};
\citet[hereafter M15]{miyaji2015}; \citet{vito2014};
\citet{buchner2015} and \citet[hereafter A15]{aird2015}.  Several
methods and models have been explored over the years; the remaining
uncertainties regard the evolution of the LF at high redshift and the
(redshift-dependent) amount of obscuration.  Further progress in such
studies requires large samples containing a sizeable number of AGN at
redshift $\gtrsim 3$ and depends on knowledge of the joint ($N_H$,$z$)
distribution (U14; M15).

A common approach to the most recent estimates of the AGN LF
(e.g.\ A10; U14; M15) is to amass a very large number of AGN from
different surveys made with different instruments, ranging from
all-sky and shallow to pencil-beam and deep. Two possible pitfalls
with that approach are that different instruments have i) different
energy responses, which  sometimes, do not even overlap, as in the case of
\textit{Swift}/BAT vs.\ \chandra\ and \xmm; and ii) different
point spread functions (PSF). In case i), biases may arise if the LFs
in different bands are different  (e.g.\ because the amount of
obscuration may evolve with redshift). In case ii), large PSFs in
medium-deep surveys (e.g. the ASCA surveys, used by A10; U14; M15) may
conceal close pairs of AGN. Furthermore, using data from very
different energy bands requires detailed spectral modelling (as done
by U14) that may introduce more uncertainties.

We adopt a different approach. We build a sample with a selection as
clean and well defined as possible. We limit ourselves to \xmm\
surveys in the 2--10 keV band so that we have the same energy response
and consistent PSFs. We focus on three surveys at different levels of
depth and area; ordered from the widest and shallowest to the
narrowest and deepest, we choose the XMM-Large scale structure survey
(XMM-LSS) \citep{chiappetti2013}, the XMM-Cosmological evolution
survey (XMM-COSMOS) \citep{xmm-cosmos}, and the XMM-\chandra\ deep
field survey (XMM-CDFS) \citep{cdfscat}. Together, these surveys
provide $\sim 3000$ objects.

Our approach to absorption corrections is to use the
\textit{Swift/BAT} spectral atlas of local AGN \citep{burlon2011}, and
the 0.5--2/2--10 keV flux ratio and the redshift of the objects from
which we build the LF. With this information, we derive a conditioned
probability distribution for the amount of absorption for each AGN. We
regard this as an improvement over A10, who did not correct for
absorption; over U14, who derived the ($N_H$,$z$) distribution from a
small subset of their sample; and over M15, who took the ($N_H$,$z$)
distribution from U14. Our approach to absorption corrections
naturally accounts for the possible increase in the fraction of
absorbed AGN from $z=0$ to $\sim 2$ that has been proposed
\citep{lafranca2005,ballantyne2006,treister2006,hasinger2008,hiroi2012,iwasawa2012}.
Moreover, by integrating the absorption distributions over the flux
ratio, we are able to test whether our data are compatible with the
($N_H$,$z$) distributions obtained by U14 and by \cite{hasinger2008}.

The most recent papers (U14; M15) estimate the LFs as
parametric fits with the maximum likelihood (ML) method. We instead
use two different methods: binned estimates and Bayesian
inference. The former provides a non-parametric representation of
the LF, which is very useful to investigate whether there is any
feature of the data that is not reproduced by the parametric
models. Bayesian inference builds on the same likelihood function of
the ML method, but provides a more accurate ascertainment of the
allowed parameter space along with theoretically sound methods to
evaluate and compare models.  Bayesian inference has already been
used to estimate LFs by A10 and \citet{fotopoulou2016}. We take this method a step
further: We use Bayesian methods to investigate whether the models
correctly reproduce the data features and the predictive
accuracy of our LF estimate.

\smallskip

We release\footnote{On the author's website:
  \url{http://www.astro.lu.se/~piero/LFTools/index.html} and on the
  source code repository:
  \url{https://github.com/piero-ranalli/LFTools}}  the code we
developed for the analysis we describe in the form of the
package  \texttt{LFTools}, which includes programmes for
absorption corrections, binned estimates, maximum-likelihood
estimates, and Bayesian inference.

In Sect.~\ref{sec:samples}, we present the samples and surveys
from which they are drawn. In Sect.~\ref{sec:nhcorr}, we introduce our
method to correct luminosities for the amount of absorption. In
Sect.~\ref{sec:binnedLF}, we derive binned estimates of the LF. In
Sect.~\ref{sec:bayes}, we introduce parametric forms for the LF, and
present methods and results from Bayesian inference.  In
Sect.~\ref{sec:modelcomparison} we introduce the concept of posterior
predictive power, and estimate this power using the Watanabe-Akaike
information criterion. In Sect.~\ref{sec:ldde-ueda14} we consider a
proposed extension to the LDDE model. In Sect.~\ref{sec:discussion} we
discuss our results.  Finally in Sect.~\ref{sec:conclusion} we present
our conclusions.

\begin{figure}
  \centering
  \resizebox{\hsize}{!}{\includegraphics[width=\columnwidth]{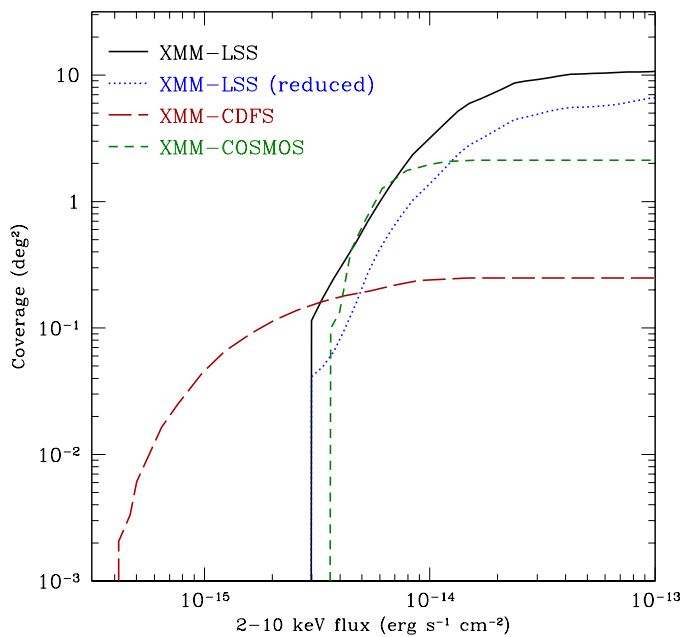}}
  \caption{ Areas covered by the XMM-LSS (black solid curve), XMM-CDFS
    (red long-dashed curve), and XMM-COSMOS (green short-dashed curve)
    surveys, vs.\ 2--10 keV flux. Although the nominal depth of LSS is
    10 ks, it also includes the Subaru Deep Field \citep{ueda2008}
    whose exposure is 100 ks.  The blue dotted curve shows the LSS
    coverage after including selection effects (availability of
    redshifts and reliability of the optical counterpart identification;
    i.e.\  $\Omega(f)$ from Eq.~(\ref{eq:reducedOmega}); see
    Sect.~\ref{sec:lssdata:fluxcorr}). The CDFS and COSMOS have nearly
    complete redshift availability (either spectroscopic of
    photometric) so selection effects can be ignored.}
  \label{fig:areas}
\end{figure}

\begin{figure}
  \centering
  \resizebox{\hsize}{!}{\includegraphics[height=\columnwidth,angle=-90,bb=73
    123 580 635,clip]{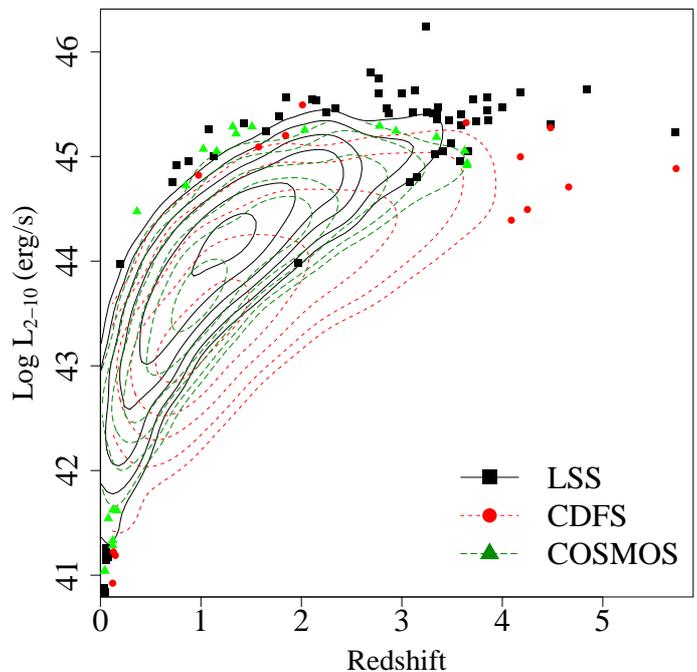}}
  \caption{ Luminosity-redshift diagram for the XMM-LSS, XMM-CDFS, and
    XMM-COSMOS surveys. The contours are logarithmically scaled and show
    the fraction of objects with $z$ and $L$ inside the contour
    (levels at 97\%, 94.8\%, 90\%, 83\%, 70\%, and 48\% from outermost
    to innermost). The data points show the objects outside the lowest
    contour.  At $z> 0.5$, the XMM-CDFS systematically probes AGN
    that are fainter by up to one order of magnitude. Observed luminosities,
    corrected only for Galactic absorption, are shown.}
  \label{fig:zlum}
\end{figure}

\smallskip

Throughout this paper, we assume a flat $\Lambda$CDM cosmology with
$H_0=70$, $\Omega_M=0.3$, and $\Omega_\Lambda=0.7$. We also assume,
for K corrections, a power-law spectrum with photon index $\Gamma=1.7$
\citep{xmmatlas}. The ``log'' and ``ln'' symbols designate the base-10
and natural logarithm, respectively.

\section{Samples}
\label{sec:samples}

The coverage curves for the XMM-LSS, XMM-COSMOS, and XMM-CDFS surveys
are shown in Fig.~\ref{fig:areas}. Two curves are shown for XMM-LSS:
the nominal coverage and a corrected coverage accounting for redshift
incompleteness (see Sect.~\ref{sec:lssdata:fluxcorr}).

The distribution in the luminosity-redshift plane of the AGN in the
XMM-LSS, XMM-COSMOS, and XMM-CDFS survey is shown in
Fig.~\ref{fig:zlum}. The peaks of the distributions occur all around
$z\sim 1.1$--1.2, but at three distinct luminosities: $L\sim
1.7\e{44}$, $4.8\e{43}$, and $2.0\e{43}$ \ergs for XMM-LSS, XMM-COSMOS, and
XMM-CDFS, respectively. The XMM-CDFS probes a
complementary part of the luminosity-redshift plane with respect to
the other two surveys, allowing one to reach luminosities which, redshifts
being equal, are one order of magnitude fainter.

\subsection{X-ray selected AGN from the XMM-LSS}
\label{sec:lssdata}

The XMM-LSS catalogue \citep{chiappetti2013} contains 2573 objects
with a hard X-ray detection and with point-like morphology. Of these,
459 have a spectroscopic redshift determination, while 1846 (among
which all the 459 are included) have a photometric redshift. The
remaining have either no optical counterpart or no photometric
redshift; this is mostly due to non-uniform optical coverage of the
field.

The X-ray catalogue contains matches to optical sources obtained with
the likelihood ratio technique, which provides a probability value for
the X-ray/optical match given by considering the sky coordinates and
their uncertainties. In this work, we only consider sources with match
probability larger than 95\%, so that our LSS sample consists of 1520
objects.
Histograms of the fluxes of the objects in the two samples are
presented in Fig.~\ref{fig:lss-sample-comparison}.
The fraction of objects used for this paper is therefore
$1520/2573\sim 59\%$.

\subsubsection*{Photometric redshifts}

Photometric redshifts (photo-$z$ in the following) for LSS
\citep{melnyk2013} were recomputed  using the LePhare SED
fitting code \citep{ilbert2006-lephare,arnouts1999-lephare}, with a
flat prior on the redshift distribution, in order to obtain the
photo-$z$ probability distributions.
We stress that we do not use
photo-$z$ at their nominal value; we rather consider for each source
its own entire probability density (see Sects.~\ref{sec:binnedLF} and \ref{sec:likelihood}).

\subsection{X-ray selected AGN from the XMM-CDFS}
\label{sec:cdfsdata}

The redshift information is almost complete (95.3\%, or 323 objects
out of 339) so no correction for incompleteness is needed.

\subsubsection*{Photometric redshifts}

The XMM-CDFS catalogue \citep{cdfscat} contains photometric redshifts
from different published sources available at the time it was
compiled. Probability distributions were however not
available. Therefore, we use photometric redshifts from
\cite{hsu2014}, so that the full probability distributions can be included.

\subsection{XMM-COSMOS}

The XMM-COSMOS catalogue \citep{xmm-cosmos} contains 1079 sources with
detection in the 2--10 keV band; 1044 of these have a redshift. Since
the redshift completeness is 97\%, completeness corrections are not
necessary. 

\subsection*{Photometric redshifts}

Probability distributions for photometric redshift are
taken from from \citet{salvato2011}, which is an update over \citet{salvato2009}.

\subsection{Corrections for incompleteness of redshift and match}
\label{sec:lssdata:fluxcorr}

\begin{figure}
  \centering
  \resizebox{\hsize}{!}{\includegraphics[width=\columnwidth]{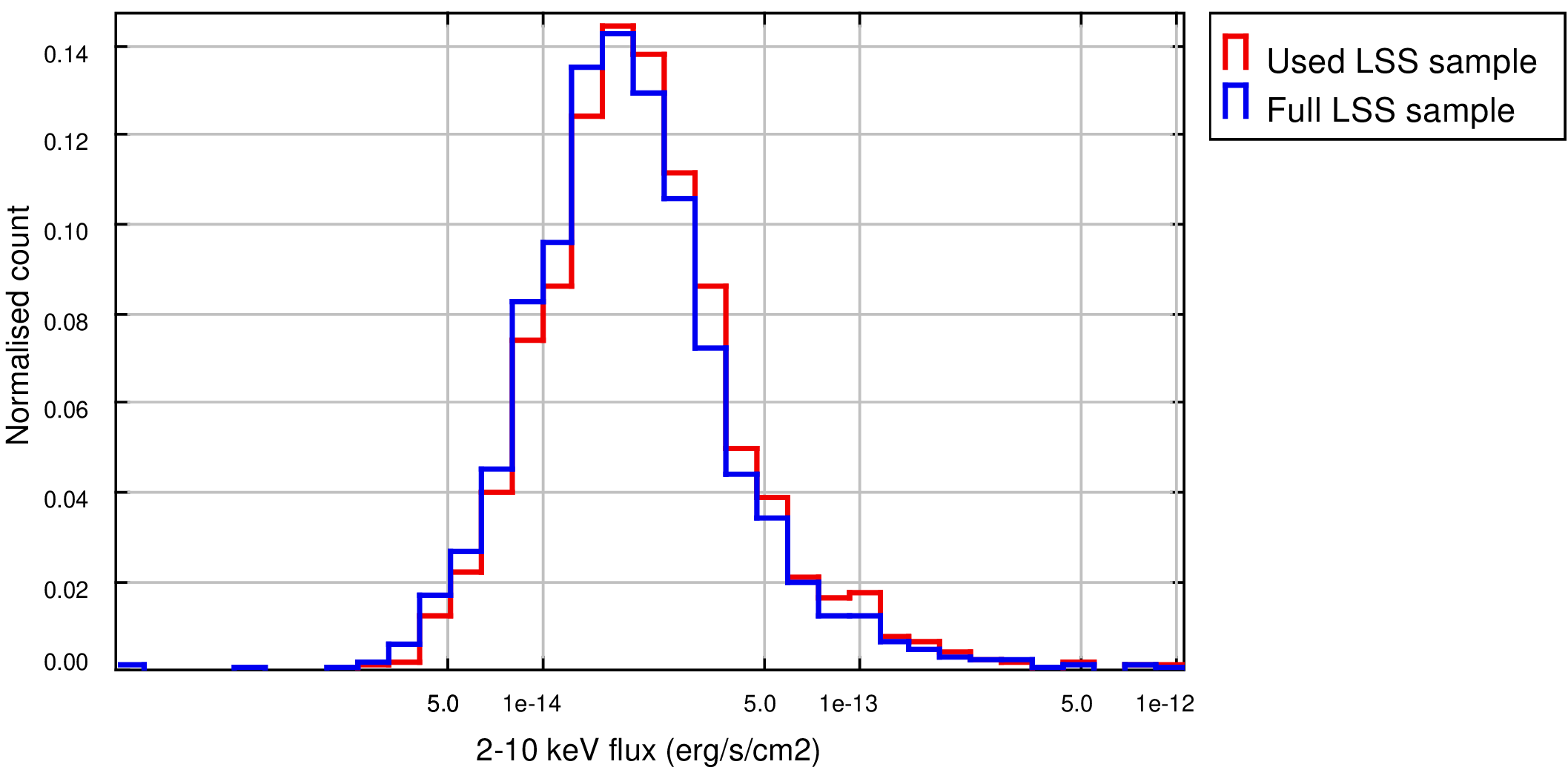}}
  \caption{ Normalised histograms of LSS fluxes. The blue histogram shows the
    full sample of 2573 objects, the red histogram shows the sample we used  that contains 1520 objects with a redshift and an optical
    counterpart match probability of $>95\%$.
  }
  \label{fig:lss-sample-comparison}
\end{figure}

The coverage should be corrected to reflect the above selection; we define
the corrected coverage $\Omega(f)$ at flux $f$ as 
\begin{equation}
\label{eq:reducedOmega}
\Omega(f) = C(f)\, \Omega_\mathrm{tot}(f)
,\end{equation}
where $C(f)$ is the correction to be made at flux $f$, and 
$\Omega_\mathrm{tot}$ is the uncorrected coverage.
We use a simple model in which $C(f)$ only
depends on the X-ray flux
\begin{equation}
\label{eq:C_f_}
C(f_i) = \frac{N(f_i)}{N_\mathrm{tot}(f_i)}
,\end{equation}
where the correction $C(f_i)$ is assumed equal to the ratio of the
number of selected objects ($N$) over the total number of objects
($N_\mathrm{tot}$) in a flux bin with centre $f_i$ and width $\Delta
\mathrm{Log}f=0.1$ (Fig.~\ref{fig:lss-sample-comparison}). The
quantities $\Omega_\mathrm{tot}$ and $\Omega$ from
Eq.~(\ref{eq:reducedOmega}) are shown in Fig.~\ref{fig:areas} as the
black solid and blue dotted curve, respectively.  When needed, $C(f)$
is interpolated over the $C(f_i)$.
The same method has been used by M15.

The above method implicitly assumes that, in each flux bin, the
objects available for the LF (i.e. those with a reliable optical
counterpart and redshift) have the same characteristics as the objects
that are not available for the LF. This assumption could be violated in some cases;
we illustrate this with an example. We consider the LF of the
objects with only a spectroscopic redshift.  There is a selection
effect that spectroscopic objects have on average lower redshifts
than objects with photometry alone. This is due to \textit{i)}
brighter objects that make spectroscopy feasible or less expensive; and
\textit{ii)} more spectral lines that are available at optical
wavelengths for low-redshift objects, while high-redshift objects often
need infrared spectroscopy, which is more demanding in terms of
instrument availability. Thus, the spectroscopic-only LF should be a
little higher at low redshifts and a little lower at high redshifts
than the spectroscopic+photometric LF. This effect can actually be
seen in our data. We found that the spectroscopic-only LF lies a
factor $\sim 1.5$ above the spectroscopic+photometric LF at
$z\lesssim 1$, and that this behaviour is reversed at $z\gtrsim
1$. This threshold at $z\sim 1$ is coherent with the redshift
distribution, which indicates that the spectroscopic $z$ is the majority at
$z\lesssim 1$ and the photometric $z$ prevails at values $\gtrsim 1$. We
stress, however, that such a factor $1.5$ is still within the
$1\sigma$ uncertainty of the binned LF presented in
Sect.~\ref{sec:binnedLF}. We consider our simple method to be
appropriate for the present data; we caution that a more articulated
treatment may be needed in case of severely incomplete, or biased,
samples.

\section{Probability distributions for absorption corrections} 
\label{sec:nhcorr}

\begin{figure*}
  \centering
  \includegraphics[width=.9\textwidth]{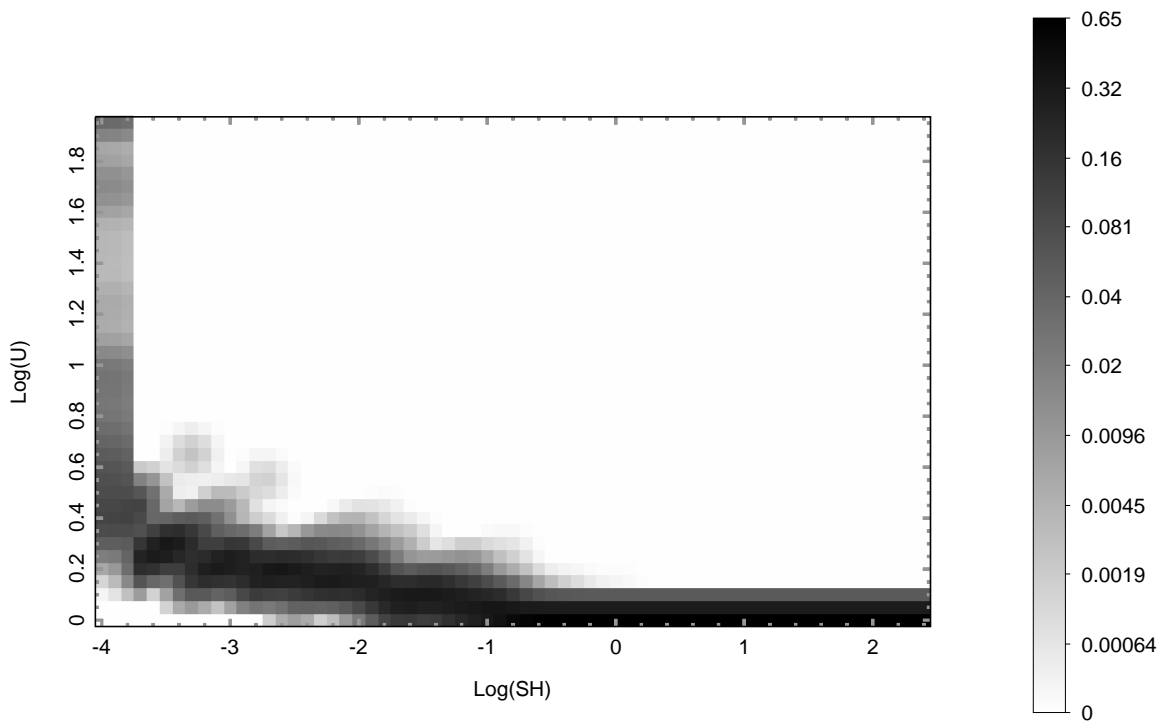}
  \caption{ Probability distribution of the unabsorbed/absorbed flux
    ratio in the 2-10 keV, as a function of the measured 0.5-2/2-10
    keV flux ratio, for the Burlon sample, at redshift zero.  The
    x-axis shows $\Log(SH)$, therefore hard sources are on the left
    and soft sources are on the right. The y-axis shows $\Log(U)$,
    therefore no correction is at the bottom and large corrections are
    at the top. The colour scale goes from white (low probability) to
    black (high probability).}
  \label{fig:nhprob}
\end{figure*}

The fluxes quoted in both the LSS and CDFS catalogues are not
corrected for intrinsic absorption. However, the LF should be computed with
intrinsic luminosities, i.e.\ corrected for absorption. Ideally,
corrections should be made by determining the column density of every
source by means of spectral fits, but this may not be feasible for
faint sources and/or for large samples. Instead, here we apply a
statistical correction based on the band ratio $SH$ between the fluxes
in the 0.5--2 and 2--10 keV bands%
\footnote{Since fluxes are computed from count rates assuming a common
  conversion factor, the distributions we describe in the following
  depend on the spectrum assumed for the catalogues. The three
  catalogues we used  all assumed a simple power law with
  $\Gamma=1.7$; we use this model accordingly. }.

The band ratio is chosen over the
hardness ratio, which would convey the same information, because
fluxes are more readily available than counts in the LSS catalogue, and because the band ratio does not depend on the instrument (while
the hardness ratio does) so that this method can also be applied to
surveys with future instruments.

Let $U$ be the ratio between the 2--10 keV unabsorbed and absorbed
fluxes.  Different combinations of column densities, spectral slope,
and redshift may lead to similar values of $SH$ (or $U$), but there is
no one-to-one correspondence between $SH$ and $U$. At any redshift, we
consider the conditional probability $P(U|SH)$, which is\ the probability
of each possible correction $U$ for a source with a given
$SH$. Following the definition of conditional probabilities we can
express it in terms of the joint probability of $U$ and $SH$,
normalised by $P(SH)$,
\begin{equation}
  \label{eq:PUSH}
  P(U|SH) = P(U \cap SH) / P(SH)\quad.
\end{equation}

To calculate the above distributions, we consider a simple spectral
model consisting of an absorbed power law, and a grid of values for
its parameters: column density $N_H$, photon index $\Gamma$, and
redshift $z$.  We obtain $U$ and $SH$ from the
model for all of the points in the grid.

An estimate of the distributions of $U$ and $SH$ at a given $z$ can
therefore be obtained by considering the joint distribution of ($N_H$,
$\Gamma$) at that $z$. For simplicity, we assume the distribution
$P(N_H, \Gamma)$ to be independent from $z$. In the following, we take
the $P(N_H, \Gamma)$ distribution from the complete sample of AGN
detected by Swift/BAT and collected by \citet{burlon2011} (hereafter
``Burlon sample'').  The Burlon sample offers good-quality spectra for
a volume-limited set of AGN. This sample has been selected in a harder energy
band than the 2--10 keV of this work, so this ensures that the sample is  much less
biased against absorbed sources, than if we used a spectral atlas selected in
the 2--10 keV band. There are several estimates of the
fraction of absorbed AGN in the literature, which we review in
Sect.~\ref{sec:CTfraction}.

With regard to the absorption correction, the main feature is the
presence of heavily-obscured objects, which produces a very wide
distribution of $U$ for sources with no 0.5--2 keV detection (leftmost
column in Fig.~\ref{fig:nhprob}); see discussion in
Sect.~\ref{sec:CTfraction}.

To smooth the distribution, we consider the parameters both at their
face values, and after adding Gaussian random errors with standard
deviations equal to those resulting from the spectral fits.

The $P(U|SH)$ distributions at $z=0,$ resulting from the Burlon sample,
are plotted in Fig.~\ref{fig:nhprob}. The main features are:
\begin{itemize}
\item as expected, softer sources are on average less absorbed and
  need fewer corrections;
\item sources without a detection in the 0.5--2 keV band (i.e.\ with
  $SH=0$) may potentially need large corrections; their $P(U)$ is
  considerably broader than that of sources with $SH>0$. This however
  depends on whether a hard $\Gamma$ or a large $N_H$ is preferred when
  fitting sources with low-quality spectra.

\end{itemize}

\begin{figure}
  \centering
  \includegraphics[width=.9\columnwidth,bb=36 182 526 608,clip]{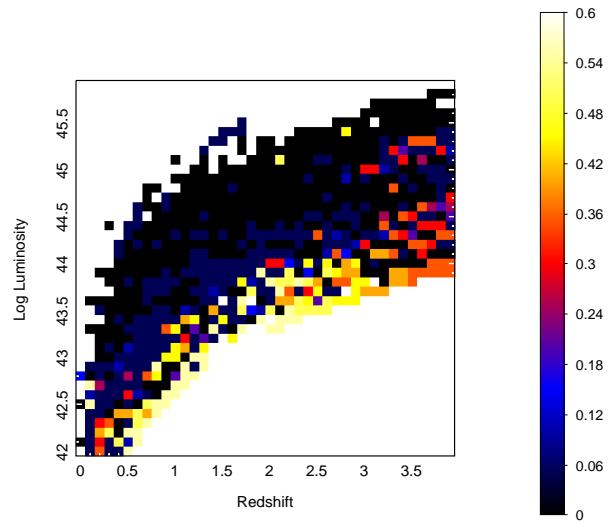}
  \caption{Medians of the absorption correction $U$ for the objects in
    the XMM-LSS, XMM-CDFS, and XMM-COSMOS surveys, over a grid of
    redshift and observed luminosity. Probability distributions for
    photometric redshifts have been included. The corrections are
    mostly significant for objects in the faint tail of the luminosity
    function (i.e. $L\lesssim 43$--$44.5$, depending on $z$), or for
    $z\gtrsim 3$. The correction $U$ (Sect.~\ref{sec:nhcorr}) is
    logarithmic, hence $U=0.5$ corresponds to the luminosity being
    corrected by a factor of 2.  The colour scale goes from black
    (median correction is 0) to bright yellow (median correction is
    0.6). White areas are not populated.  }
  \label{fig:ulz}
\end{figure}

In Fig.~\ref{fig:ulz}, we show the median correction $U,$ which resulted
from applying our method to the surveys described in this paper. The
objects were grouped according to their redshift and observed
luminosity (considering probability distributions of photometric
redshift).  For $z\lesssim 3$, corrections are small for objects in
the bright tail of the LF, and larger for objects in the faint
tail. At $z\gtrsim 3$, larger corrections also appear for
high-luminosity objects ($L>10^{44.5}$). 

\subsection{Fraction of heavily obscured AGN}
\label{sec:CTfraction}

In this Sect., we compare several estimates of the fraction of
obscured ($N_H>2\e{22}$ \cmq) and Compton-thick ($N_H>1\e{24}$ \cmq;
hereafter CT) AGN in the literature. Our aim is to check that our use
of the \cite{burlon2011} sample is consistent with current knowledge.

\citet{burlon2011} analysed a sample of 199
spectra of AGN selected in the 15--200 keV band with
\textit{Swift}/BAT, finding that 53\% were obscured and 5.5\%
CT\footnote{\citet{burlon2011} report a fraction of 4.6\% but they
  define CT as having $N_H>1.5\e{24}$ \cmq.}. They also claim that
after correcting for the observational bias, which makes CT sources
difficult to detect, the CT fraction could rise to $20^{+9}_{-6}\%$.
Similar fractions, of 7\% CT and 43\% obscured AGN, have been found by
\citet{malizia2009} in an INTEGRAL-selected complete sample.

At lower energies, \citet{brightman2012} reanalysed \chandra\ spectra
in the CDFS using models accounting for Compton scattering and the
geometry of circumnuclear material, finding a fraction of 5.5\% CT
objects. These authors estimate that after accounting for the observational
bias, this fraction should rise to $\sim 20\%$ in the local universe,
and to $\sim 40\%$ at $z=1$--4.
In a sample of galaxies selected in the infrared with magnitude $K<22$
and with $1.4<z<2.5$, \citet{daddi2007} found a fraction of 20\% CT.

From fits to the cosmic X-ray background, \citet{akylas2012} found
that a fraction of 5--50\% is allowed. Fractions of the order of 10\%
are also reported in \citet{treister2009} and in the models by
\citet{hopkins2006cxb} and \citet{gilli2007}.  At high redshift and
for intermediate luminosities, \cite{hasinger2008} reports that there
is ``convincing evidence'' that there is no large change in the
relative numbers of Compton-thin and -thick AGN with respect to the
local universe.

Our use of the \citet{burlon2011} sample to derive the absorption
correction is therefore supported by the available literature. If the
larger fractions of CT AGN obtained after correcting for the
observational bias were held true, then our corrections could be
regarded as very conservative.

\subsection{High-redshift evolution of the absorbed fraction}
\label{sec:high-z-nh}

The fraction of absorbed AGN seems to be larger at high redshift than
in the local universe
\citep{lafranca2005,ballantyne2006,treister2006}. \citet{hasinger2008}
found an increase that could be modelled as $(1+z)^{0.62\pm 0.11}$
for $0<z<2$, saturating at $z\sim 2$. This is approximately a factor
of 2 at $z\sim 2$. \cite{ueda2014} found an increase by a factor of
$\sim 1.5$ between the redshift intervals $0.1<z<1$ and $1<z<3$.  

The \cite{burlon2011} sample consists of objects at $z\sim 0$, so one
could ask if the absorption evolution could have any effect on our
method.  We never use the marginal distribution $P(N_H)$ from
\citet{burlon2011}; instead in Eq.~(\ref{eq:PUSH}), we  use the
($N_H,\Gamma$) joint distribution to obtain the conditioned
probability of the needed correction, given the observed soft/hard
flux ratio. The absorption correction mostly depends on $N_H$,
thus $P(U|SH)$ is essentially analogous to $P(N_H|SH)$. The evolution
of absorption is reflected in an evolution of the flux ratio $SH$,
therefore our method naturally accounts for the evolution of
absorption.  

One possibility is that the \citet{burlon2011} sample is still missing
some kind of $(N_H,\Gamma)$ combination, which is rare in the local
universe but  becomes abundant at high redshift (or, conversely,
something that is abundant becomes rare). For example, if
Compton-thick AGN, were more abundant at high redshift, then our
corrections would err on the conservative side (less correction than
needed). We prefer not to speculate on how the Compton-thick
population (or that of any other kind of AGN) changes with
redshift. However, we stress that \textit{i)} the \cite{burlon2011}
sample contains 11 Compton-thick AGN (6\% of total) so our corrections
are not over-influenced by just one or a few objects; and \textit{ii)}
the fraction of Compton-thick AGN does not seem to vary much with
redshift (see Sect.~\ref{sec:CTfraction}).

\section{Binned luminosity function}
\label{sec:binnedLF}

The differential luminosity function $\Phi$ is defined as the number
of objects $N$ per comoving volume $V$ and per unabsorbed luminosity $L$ as follows:
\begin{equation}
\Phi(L,z) = \frac{ \de^2 N(L,z) } { \de V\,\de L }
.\end{equation}

A few variants of the original $1/V_\mathrm{max}$ method
\citep{schmidt68} have been proposed with the aim of refining the
method; examples are \citet{pageca00,lafranca1997,miyaji2001}.  Here
we build on \citet{pageca00} and include absorption corrections and
probability distributions for photometric redshift in the method.

\begin{figure*}
  \centering
  \includegraphics[height=.9\textwidth,angle=-90]{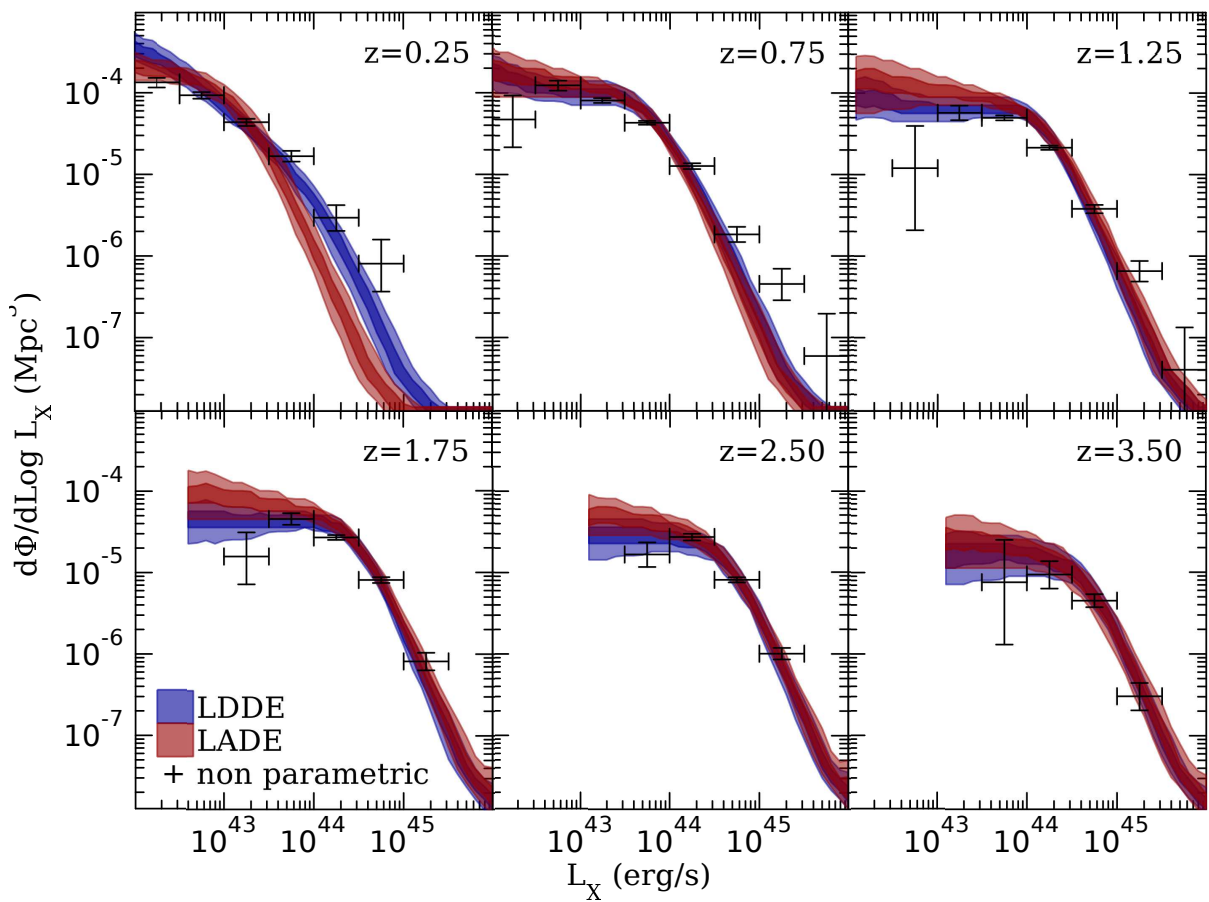}
  \caption{ Luminosity function, from combined XMM-LSS, CDFS, and
    COSMOS data, with binned estimates (black data points with
    $1\sigma$ error bars) and Bayesian highest posterior densities
    (HPD) under the LDDE (blue area) and LADE (red area) models.  For
    both models, the darker areas show the 68.3\% HPD intervals, while
    the lighter areas show the 99.7\% HPD intervals.}
  \label{fig:binnedLF}
\end{figure*}

The LF in a bin with luminosity and redshift boundaries
$L_\mathrm{min}$, $L_\mathrm{max}$ and $z_\mathrm{min}$,
$z_\mathrm{max}$, respectively, and containing $N$ objects, is
approximated by
\begin{equation}
\label{eq:binnedLF}
\Phi(\left<L\right>,\left<z\right>) \sim 
\frac{N}{V_\mathrm{probed}}
\end{equation}
with
\begin{equation}\label{eq:vprobed}
V_\mathrm{probed} =
\int_{L_\mathrm{min}}^{L_\mathrm{max}}
\int_{z_\mathrm{min}}^{z_\mathrm{max}}
\Omega(L,z) \frac{ \de V }{ \de z }\, \de z\, \de L
,\end{equation}
where $\left<L\right>$ and $\left<z\right>$ are the log-average
luminosity and average redshift of the bin, respectively;
$\Omega(L,z)$ is the survey coverage at the flux that an object of
luminosity $L$ would have if placed at redshift $z$; and $\de V/\de z$
is the comoving volume.

For each source $i$, the unabsorbed 2--10 keV luminosity $L_i$ is obtained
from the observed 2--10 keV flux $f_i$ and the 0.5--2/2--10 keV flux
ratio $SH_i$, considering the following redshift and absorption probability
distributions (Sect.~\ref{sec:nhcorr}):
\begin{equation}
L_i(U,z)\, \de U\, \de z = f_i\,4\pi D^2(z) U P(U|SH_i,z) P(z) \,\de U\, \de z
,\end{equation}
where $D(z)$ is the luminosity distance. 
Therefore we replace $N$ in
Eq.~(\ref{eq:binnedLF}) with
\begin{equation}
\label{eq:binnedLF2}
 \sum_i
\int_{z_\mathrm{min}}^{z_\mathrm{max}}
\int_{1}^{U_\mathrm{max}}
q^{L_i} P_{i}(z)\,P(U|SH_i,z)
\,\de U\,\de z\
,\end{equation}
where $q^{L_i}$ is 1 if $L_\mathrm{min}\le L_i(U,z) < L_\mathrm{max}$,
and is 0 otherwise.  For sources with photo-$z$, $P_i(z)$ is the probability
density obtained from the template fitting, while for sources with
spectroscopic redshift, we use Dirac's $\delta$, 
$ P_i(z)=\delta(z_\mathrm{spec})$. Errors on $N$ are estimated assuming
Gaussianity (for $N\ge 50$) or by interpolating the tables in
\citet{gehrels86} (for $N<50$).

As for $V_\mathrm{probed}$ (Eq.~\ref{eq:vprobed}), the integrals should
run on the unabsorbed luminosities, while the coverage $\Omega(L,z)$
should refer to the observed (i.e.\ absorbed) fluxes.  Therefore we
replace Eq.~(\ref{eq:vprobed}) with
\begin{equation}\label{eq:vprobed-corr}
V_\mathrm{probed} =
\int\limits_{L_\mathrm{min}}^{L_\mathrm{max}}
\int\limits_{z_\mathrm{min}}^{z_\mathrm{max}}
\int\limits_0^{U_\mathrm{max}}
\Omega\left( \frac{L}{U},z \right) P(U|z)\, \de U\,  \frac{ \de V }{ \de z }\, \de z\, \de L
,\end{equation}
where $L/U$ is therefore the absorbed luminosity, and $P(U|z)$ is the
marginal probability of an absorption correction $U$, conditioned only
by the redshift $z$.

The binned LF is shown in Fig.~\ref{fig:binnedLF} (and also, for
comparison, in Figs.~\ref{fig:LDDELFs} and \ref{fig:LADELFs}).

\section{Bayesian inference}
\label{sec:bayes}

Bayesian inference relies on a different interpretation of what a
probability is with respect to the classical (frequentist)
interpretation. In the Bayesian framework, probability measures our
degree of belief about a proposition; a probability can be assigned to
the parameters subject of inference or to abstract ideas such as
models. We will not go over the theory details here; a nice
introduction can be found in \citet{trotta2008} or in the book by
\citet{gregory2005bayesian}. Advanced methods can be found in
\citet{BDA3}. Example applications pertaining to astronomical
catalogues, source detection, flux estimate, etc. can be found in
\citet{andreon2011} and \citet{andreon2013}.

The outcome of Bayesian inference is a posterior probability
distribution that yields the
probability $\prob(\theta)$ for a vector of parameters $\theta$. There are some similarities with
maximum-likelihood (ML) methods: for example, the same likelihood
function is used. However, while ML aims to find just the best-fit
values for $\theta$ (with confidence intervals derived by asymptotic
theory), Bayesian inference aims to obtain $\prob(\theta)$ for all
possible or reasonable values of $\theta,$ and therefore offers a more
accurate description of how the model fits the data.

Exploring the parameter space becomes computationally intensive as
soon as the dimensionality of the parameters $\theta$ becomes larger
than a few; the models considered in this paper have either eight (LADE)
or nine (LDDE) dimensions. Effective methods are therefore valuable.
Nested sampling has been proposed as a particularly powerful method
\citep{skilling2004,skilling2006}, and has already been used for LF estimates
by A10. The most popular implementation is the MultiNest library
\citep{feroz2008multinest,feroz2009multinest,feroz2013multinest},
which we used to derive the posterior distributions for the LF parameters.

In this section, we first describe the ingredients needed for Bayesian
inference: parametric models of LF with redshift evolution and the
likelihood function. Next, we present our results.

\subsection{Parametric form for the luminosity function}
\label{sec:parametricLF}

A broken power-law form has been suggested for the $z\sim 0$ AGN
luminosity function since  early works
\citep{maccacaro1983,maccacaro1984} as follows:
\begin{equation}
  \label{eq:doublepow}
  \frac{\de \Phi (L)}{\de \Log L} = A \left[
    \left( \frac{L}{L_*} \right)^{\gamma_1}
  + \left( \frac{L}{L_*} \right)^{\gamma_2}
\right]^{-1}
,\end{equation}
where $A$ is the normalisation, $L_*$ is the knee luminosity, and
$\gamma_1$ and $\gamma_2$ are the slopes of the power law below and
above $L_*$.

The LF parameters however evolve with redshift
\citep{boyle1994,page1996,jones1997}. Two simple and alternative forms
for evolution are pure luminosity evolution
\citep[PLE;][]{mathez1978-ple,braccesi1980} and pure density evolution
\citep[PDE;][]{schmidt68,schmidt1983}. The basic ideas of these forms are that
$L_*$ is brighter at higher $z$ (PLE) or $A$ is larger at higher $z$
(PDE).  Several models have been proposed in the literature that
bridge between these two possibilities and provide reasonable
descriptions of the data. We  focus on the two  models that are currently most commonly used. Their complex functional forms
are justified by the necessity to allow the bright end of the LF to
move at larger luminosity at increasing $z$ (to the right, in any
panel of Fig.~\ref{fig:binnedLF}), while at the same time having the
faint end of the LF move at lower number densities (to the bottom, in
the same figure). Thus they can model the AGN downsizing: moving from
the high-redshift universe to present, the more luminous AGN have
became much fainter and the less luminous AGNs have become more common.

\subsection{Luminosity and density evolution}
The luminosity and density evolution model (LADE; \citealt{ueda03}, A10)
joins both kinds of evolution, and also enables a change in the pace of
luminosity evolution after a critical redshift $z_c$. Following
A10, we use a double power law for the luminosity
evolution
\begin{equation}
\label{eq:lade1}
  \frac{\de \Phi (L,z)}{\de \Log L} = 
    \frac{\de \Phi (L\times\eta_l(z), z=0)}{\de \Log L} \eta_d(z)
\end{equation}
with
\begin{equation}
  \label{eq:lade2}
  \eta_l(z) = \frac{1}{k} \left[
           \left( \frac{1+z_c}{1+z} \right) ^{p_1}
           +
           \left( \frac{1+z_c}{1+z} \right) ^{p_2}
           \right] \quad,
\end{equation}
\begin{equation}
  \label{eq:lade3}
  \eta_d(z) = 10^{d(1+z)}
\end{equation}
\begin{equation}
  \label{eq:lade4}
  k =      \left( 1+z_c \right) ^{p_1}
           +
           \left( 1+z_c \right) ^{p_2}
,\end{equation}
where the evolution parameters are the critical redshift $z_c$, the
luminosity evolution exponents $p_1$ and $p_2$, and the density
evolution exponent $d$. In particular, $d$ is assumed to be negative
to allow the faint end of the LF to decrease at larger $z$.

Following \citet{fotopoulou2016} and at variance with
A10, we have normalised the LADE model so that at $z=0$,
$\eta_l=1$, and Eq.~(\ref{eq:lade1}) reduces to the local LF.

\subsection{Luminosity-dependent density evolution}
Luminosity-dependent density evolution (LDDE), in the formalism
introduced by \citet{ueda03}, can be expressed as\begin{equation}
\label{eq:ldde1}
\frac{\de \Phi (L,z)}{\de \Log L} = 
    \frac{\de \Phi (L,z=0)}{\de \Log L} \times \mathrm{LDDE}(L,z)
\end{equation}
with
\begin{equation}
\label{eq:ldde2}
\mathrm{LDDE}(L,z) = \left\{ \begin{array}{ll}
  (1+z)^{p_1}                                       &z \le z_0(L)  \\
  (1+z_0)^{p_1}
     \left(\frac{1+z}{1+z_0}\right)^{p_2}\quad     &z > z_0(L)  
\end{array}  \right.
\end{equation}
and
\begin{equation}
\label{eq:ldde3}
z_0(L) = \left\{ \begin{array}{ll}
    z_c                                                & L \ge L_\alpha \\
    z_c \left( \frac{L}{L_\alpha} \right)^\alpha \qquad  & L < L_\alpha \quad .
  \end{array} \right.
\end{equation}
The evolution parameters are the critical redshift $z_c$, the
evolution exponents $p_1$ and $p_2$, and two parameters ($\alpha$ and
$L_\alpha$) that give a luminosity dependence to $z_c$. Although the
functional form is different, the main features are the same as for LADE.
The main difference is that, for increasing $z$, the slope of the
faint end of the LF stays constant in LADE, while it changes (it
flattens) in LDDE.

\smallskip
In U14, this model is further extended to include a luminosity
dependence on $p_1$ and a second break at a redshift $z>z_0$, adding
six more parameters (with a total of 15 parameters, we refer to this
extension as LDDE15). Several parameters in U14 are fixed at values,
which make the LF decline faster beyond $z\sim 3$, reproducing the
results by \citet{fiore2012}. In the following, we initially consider
the nine-parameter LDDE and compare it to LADE, deferring our treatment
of LDDE15 to Sect.~\ref{sec:ldde-ueda14}.

\subsection{Likelihood function}
\label{sec:likelihood}

The likelihood function can be obtained, following
\citet{marshall1983} \citep[see also][]{loredo2004}, by considering a
Poissonian distribution for the probability of detecting a number
$y_i$ of AGN of
given luminosity $L_i$ and redshift $z_i$,
\begin{equation}
P=\frac{(\lambda_i)^{y_i} e^{-\lambda_i}}{y_i!}
\end{equation}
with
\begin{equation}
\label{eq:lambda}
\lambda_i=\lambda(L_i,z_i)=\Phi(L_i,z_i)\, \Omega(L_i,z_i) \frac{\de V}{\de z}
\de z \,\de Log L
,\end{equation}
where $\lambda$ is the expected number of AGN with given $L_i$ and $z_i$; and
$\Phi$ is the LF evaluated at the source luminosity and redshift.

The likelihood $\mathcal{L}$ is then defined as the product of the probability of
detecting every source $i$ in the catalogue times the probability of
not detecting any AGN in the remaining parameter space $L_j$, $z_j$, i.e.,
\begin{equation}
\mathcal{L} = \prod_i \lambda(L_i,z_i) e^{-\lambda(L_i,z_i)}
\prod_j e^{-\lambda(L_j,z_j)} \quad .
\end{equation}
The product of the exponential terms is
actually extended over  the entire parameter space. Therefore, 
the log-likelihood $S=\mathrm{ln}\, \mathcal{L}$
may be written as
\begin{equation}
\label{eq:loglikelihood}
S =  \sum_i \mathrm{ln}\, \lambda(L_i,z_i) - \int\!\!\! \int \lambda(L,z) \de z \,\de Log L
\end{equation}
so that $S$ may considered as the sum of a `source term' (the
left term, which is a sum over all sources) and a `coverage
  term' (the right term, i.e.\ the integral of $\lambda$).
The integrals extend over the 0.0001--4 and $10^{41}$--$10^{46}$
\ergs\ ranges in redshift and luminosity, respectively. Also, only the
sources falling in these ranges are considered.

\citet{marshall1983} drop the coverage and comoving volume from the
sums in the first term of Eq.~(\ref{eq:loglikelihood}) because they do
not depend on the fit parameters and can be treated as constants.
Therefore, these authors obtain the same expression found by \citet{loredo2004}
using a more correct approach. Here, consistent with
\citet{loredo2004}, we drop only the coverage and keep the comoving
volume because we need to consider the dependence of $\lambda$ on the
redshift (whenever the source has a photo-$z$) and on the
absorption. Therefore we rewrite Eq.~(\ref{eq:loglikelihood}) as
\begin{equation}
\label{eq:loglikelihood2}
S =  \sum_i \mathrm{ln} \left(
   \Phi(L_i,z_i) \frac{\de V}{\de z} \right)
  - \int\!\!\! \int \lambda(L,z) \de z \,\de Log L \quad .
\end{equation}

For each source,  $\Phi_i$ is averaged over the redshift
and absorption distributions
\begin{equation}
\begin{array}{l}
\label{eq:loglikelihood3}
\Phi(L_i,z_i) \frac{\de V}{\de z} = \\
\\
   \quad \int\!\!\! \int \Phi(L(f_i,z,U),z) \frac{\de V}{\de z} P_i(z)
   P(U|SH_i,z) \de U \de z \quad .
\\
\end{array}
\end{equation}
For sources with a spectroscopic redshift, $P(z)$ may be interpreted
as a $\delta$ distribution as carried out for the binned LF.

As noticed for the binned LF, the survey coverage should refer to
observed, i.e.\ absorbed, fluxes. Therefore in the coverage
term in Eq.~(\ref{eq:loglikelihood2}), $\lambda$ should be (compare
with Eq.~\ref{eq:vprobed-corr} and Eq.~\ref{eq:lambda})
\begin{equation}
\lambda(L,z) = \Phi(L,z) \frac{\de V}{\de z}
\int \Omega\left( \frac{L}{U} \right) P(U|z) \, \de U \quad.
\end{equation}

Selection effects can be included in the likelihood function
(Eq.~\ref{eq:loglikelihood2}) following A10. If the expected
number of objects $\lambda$ is reduced by the factor $C(L,z)$ defined
in Eq.~(\ref{eq:C_f_}) as
\begin{equation}
\lambda^\prime(L,z) = C(L,z)\, \lambda(L,z)
,\end{equation}
then $\lambda$ in Eq.~(\ref{eq:loglikelihood2}) has to be replaced by
$\lambda^\prime$. 
In practice, this amounts to the reduction of the
coverage function introduced in Eq.~(\ref{eq:reducedOmega}).

\begin{figure*}[p]
  \centering
  \includegraphics[height=.8\textwidth,angle=-90]{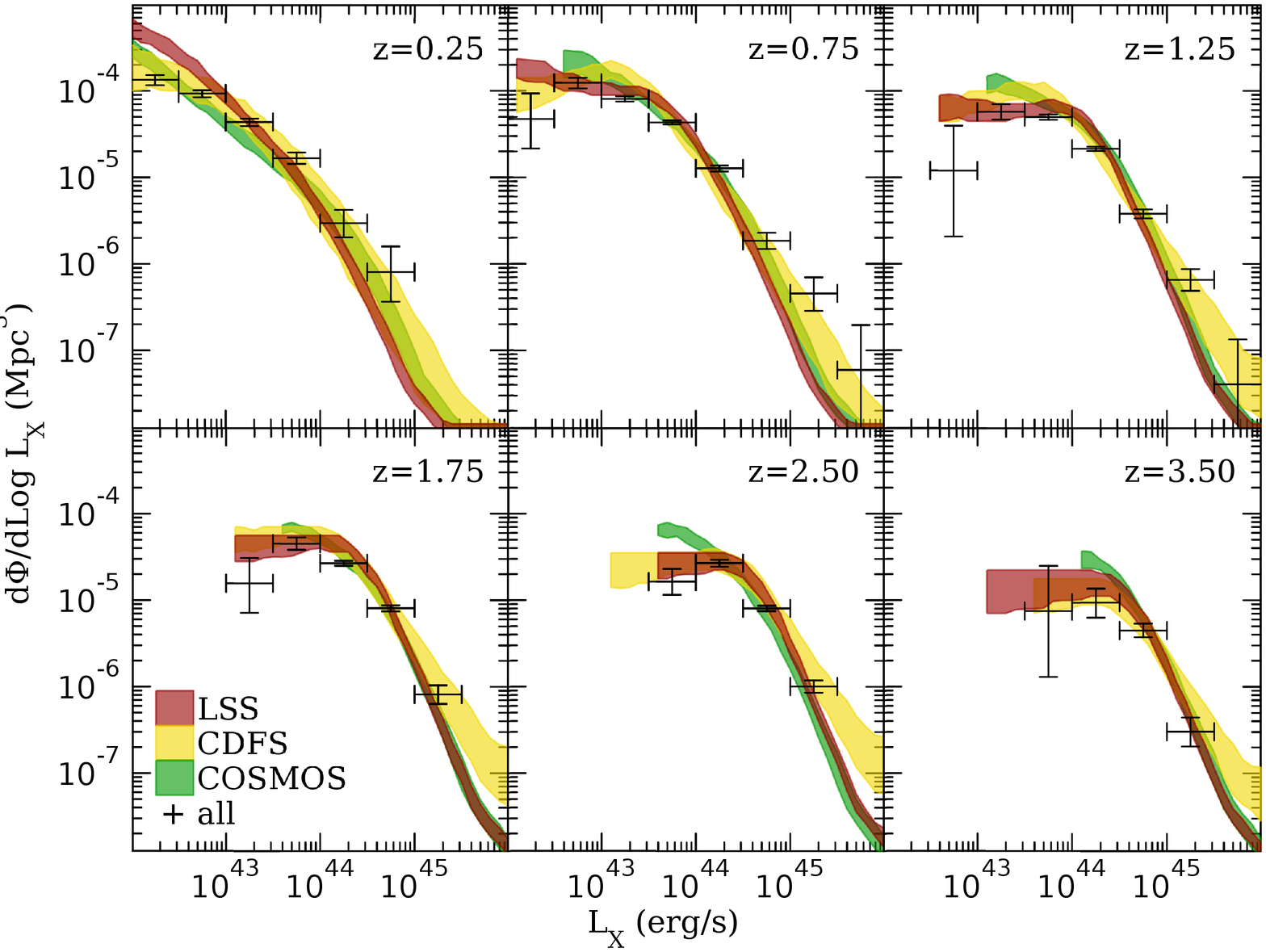}
  \caption{ Luminosity function, from XMM-LSS, CDFS, and COSMOS
    data, with binned estimates from all surveys together (black data
    points) and Bayesian highest posterior densities (68.3\% HPD interval) for individual
    surveys under the LDDE model (red: LSS; yellow: CDFS; green: COSMOS).}
  \label{fig:LDDELFs}
\end{figure*}

\begin{figure*}[p]
  \centering
  \includegraphics[height=.8\textwidth,angle=-90]{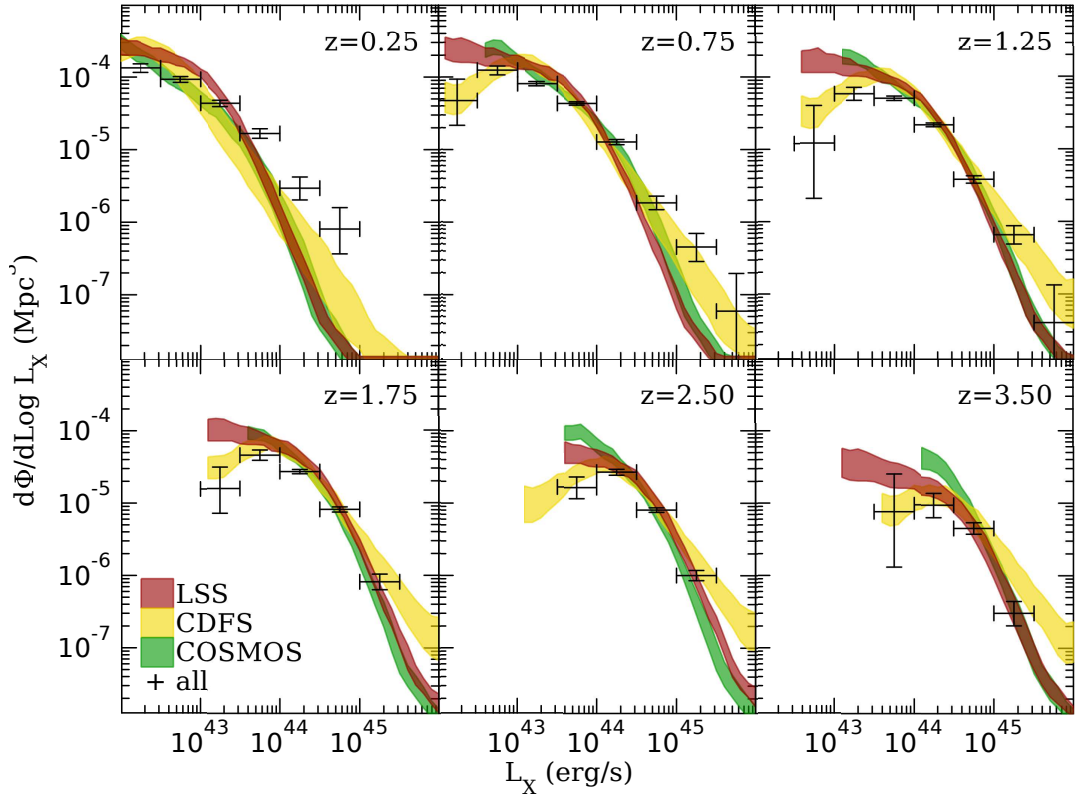}
  \caption{ Luminosity function, from XMM-LSS, CDFS, and COSMOS
    data, with binned estimates from all surveys together (black data
    points) and Bayesian highest posterior densities (68.3\% HPD interval) for individual
    surveys under the LADE model.  Colours
    as in Fig.~\ref{fig:LDDELFs}.}
  \label{fig:LADELFs}
\end{figure*}

\begin{figure*}
  \centering
  \includegraphics[height=.9\textwidth,angle=-90]{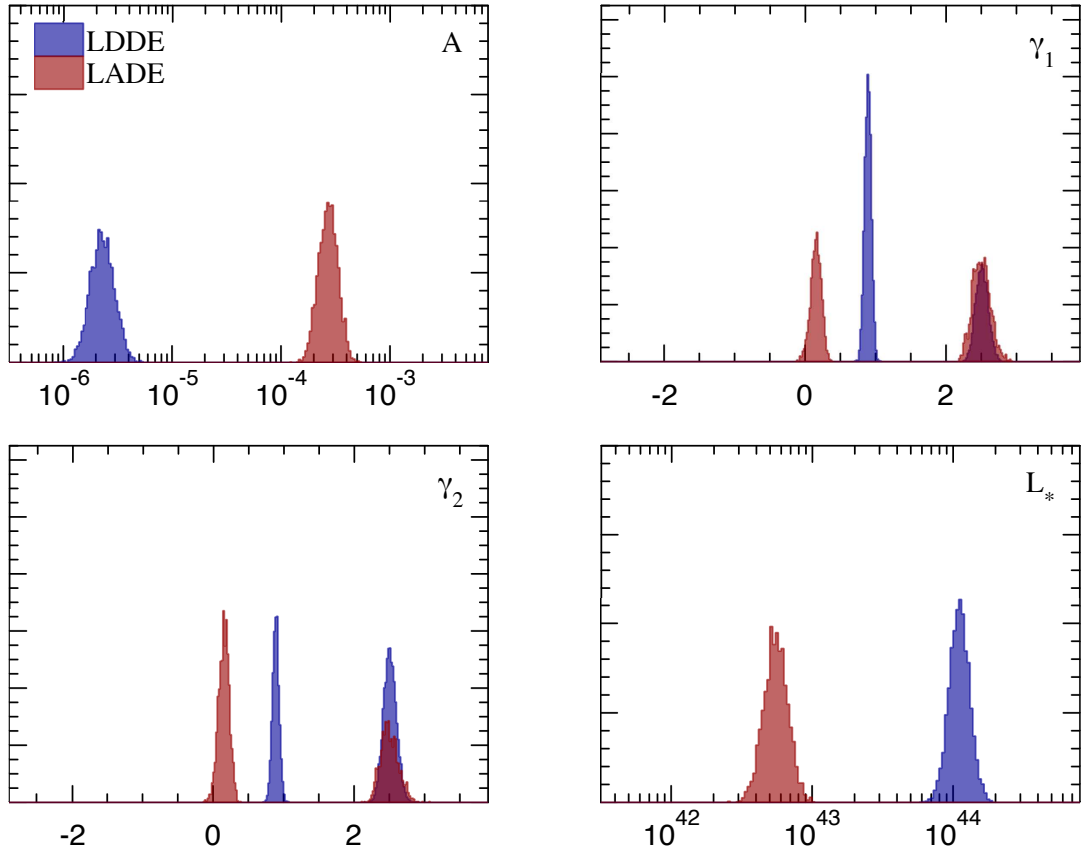}
  \caption{ Posterior probability densities for double power-law
    parameters for all surveys combined, under the LDDE (blue) and
    LADE (red) models. The plotted variables are identified in the top
    right corner of each panel. The normalisation $A$ is in
    Mpc$^{-3}$, the knee luminosity $L_*$ is in \ergs. The histograms
    are normalised hence the vertical scales are arbitrary.}
  \label{fig:postmodels}
\end{figure*}

\begin{figure*}
  \centering
  \includegraphics[height=.9\textwidth,angle=-90]{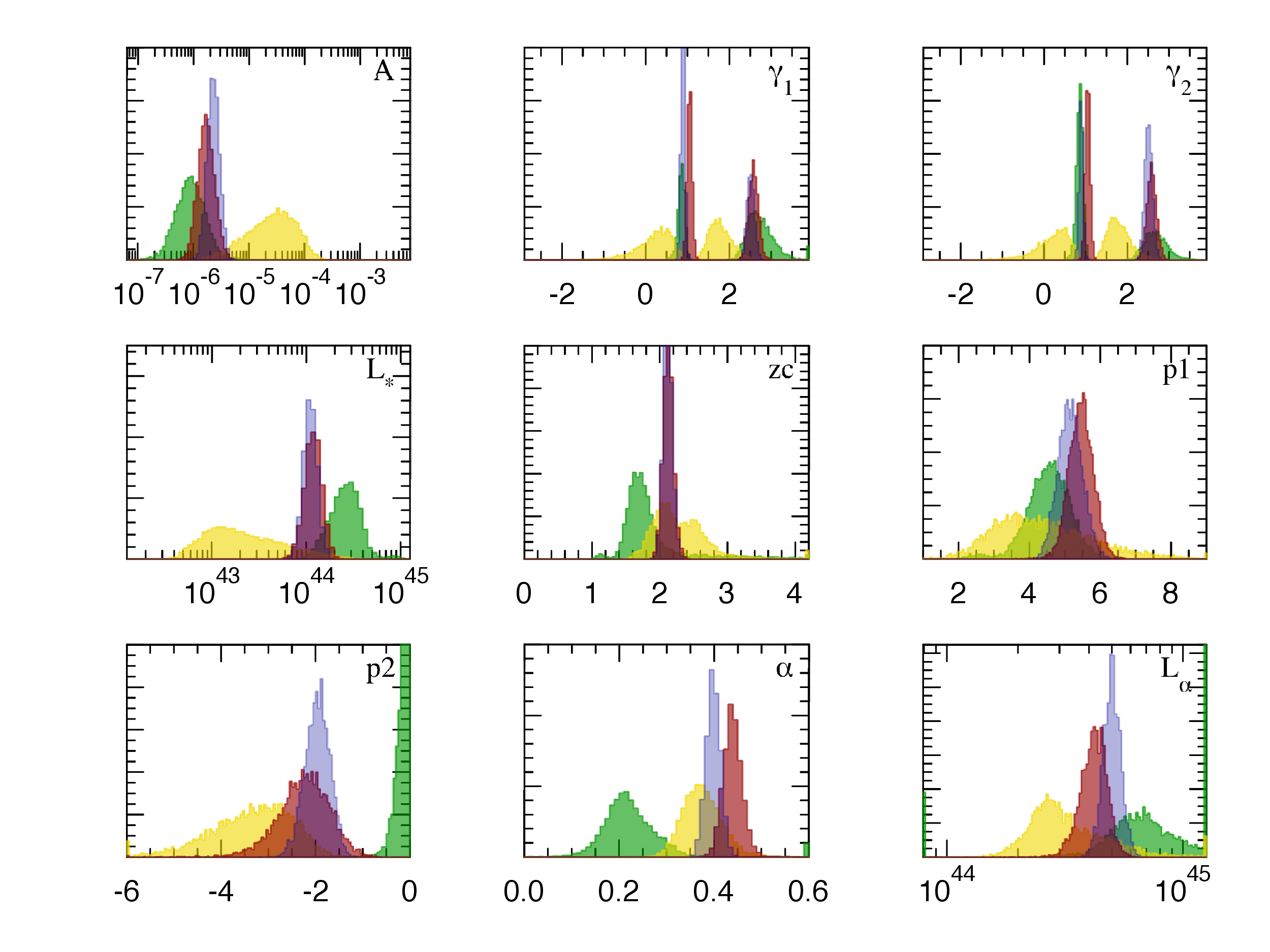}
  \caption{ Posterior probability densities for double power-law and
    evolution parameters for individual surveys under the LDDE model;
    colours as in Fig.~\ref{fig:LDDELFs}. For comparison, we also plot
    (blue histogram) the probability density for all surveys together
    under the same model. The plotted variables are identified in the
    top right corner of each panel. The normalisation $A$ is in
    Mpc$^{-3}$, the luminosities $L_*$ and $L_\alpha$ are in
    \ergs. The histograms are normalised hence the vertical scales are
    arbitrary.}
  \label{fig:postLDDE}
\end{figure*}

\begin{figure*}
  \centering
  \includegraphics[height=.9\textwidth,angle=-90]{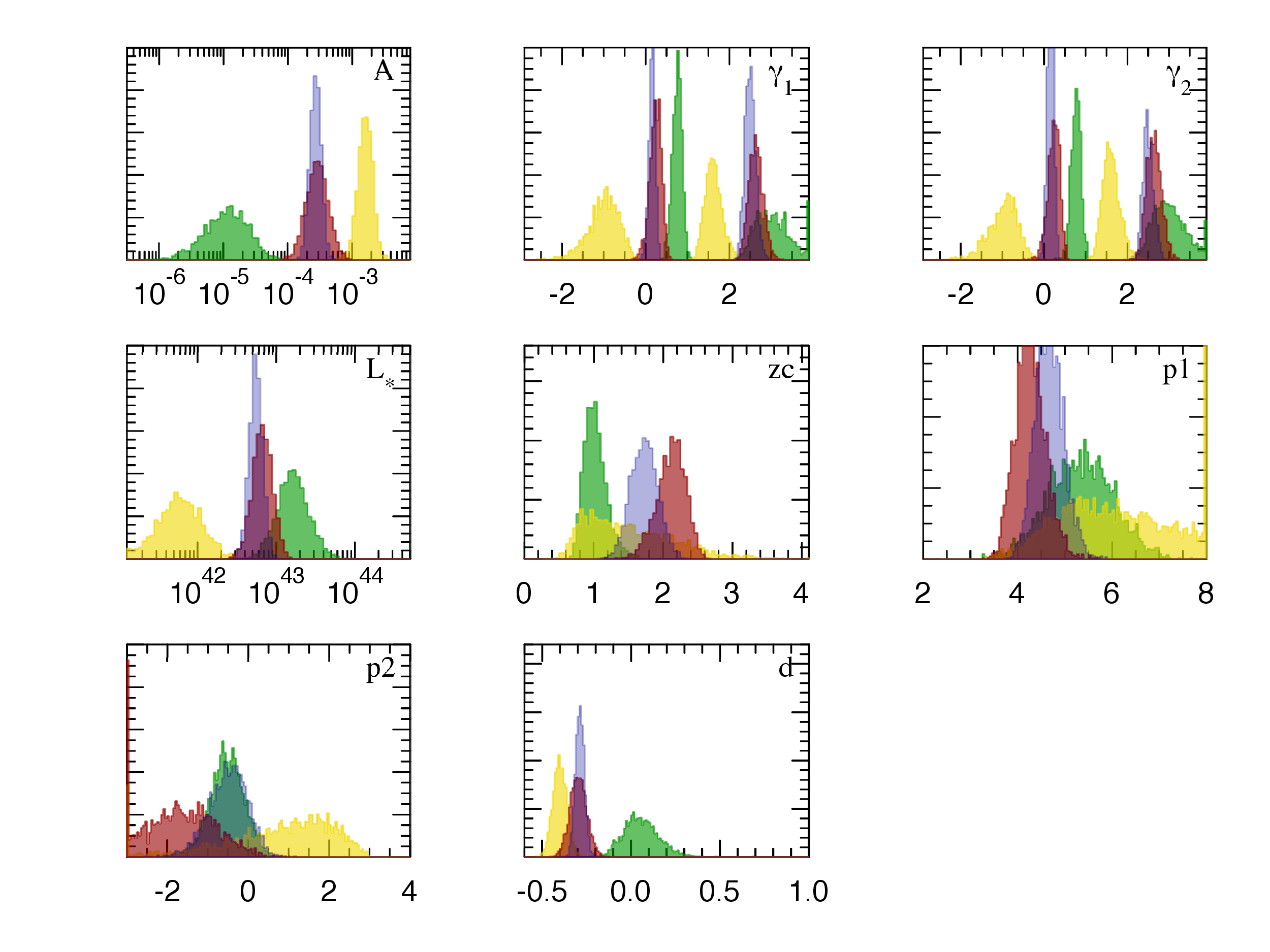}
  \caption{ Posterior probability densities for double power-law
    parameters for individual surveys under the LADE model; colours
    as in Fig.~\ref{fig:LDDELFs}. For comparison, we also plot (blue
    histogram) the probability density for all surveys together under
    the same model. The plotted variables are identified in the top
    right corner of each panel. The normalisation $A$ is in
    Mpc$^{-3}$, the knee luminosity $L_*$ is in \ergs. The histograms
    are normalised hence the vertical scales are arbitrary.}
  \label{fig:postLADE}
\end{figure*}

\subsection{Results}
\label{sec:bayesianresults}

We used the nested sampling method, together with the likelihood
function and the parametric form described above, to compute the
posterior probability distribution (hereafter just posterior) for the
LF parameters. We repeated the computation for four combinations of
data (all surveys together, and each survey individually) and models
(LADE and LDDE).  The result of each computation is a set of
$\sim 7000$--8000 draws from the posterior; the exact number depends
on the individual run, and on when the MultiNest algorithm attains
convergence. These are available as online-only
material.
In Tables~\ref{tab:HPD-LADE} and \ref{tab:HPD-LDDE} we summarise the
content of the draws: for each parameter, we report the
mode\footnote{Formally, the mode is ill-defined for a sample of
  floating point numbers. In practice, the location of the peak of a
  histogram of the drawn values gives the most probable value; this is
  what we report. The mode of the posterior is usually suggested as a
  Bayesian analogue to the best-fit value in frequentist statistics.}
and the 68.3\% (``$1\sigma$'') highest posterior density (HPD)
interval\footnote{The number of $\sigma$s is put between quotes to
  remember that posterior distributions are not necessarily Gaussian,
  hence speaking of $\sigma$ is not formally proper.}. The HPD is the
interval containing a given fraction of the posterior density,
such that the posterior density inside the HPD is always larger than
outside.

\begin{table}[t]
  \centering
  \begin{tabular}{lrrr}
    \hline\hline
    Parameter & Mode    & \multicolumn{2}{c}{68.3\% HPD interval}   \\
              &         & $\qquad$min & max     \\
    \hline
$\Log\,A$     & -3.53    & -3.65        & -3.48    \\
$\gamma_1$    & 0.16    & 0.09        & 0.23    \\
$\gamma_2$    & 2.48    & 2.37        & 2.60    \\
$\Log\, L_*$  & 42.72   & 42.65       & 42.82   \\
$z_c$         & 1.72    & 1.53        & 1.93    \\
$p_1$         & 4.67    & 4.35        & 5.00    \\
$p_2$         & $-0.30$ & $-0.91$     & $0.02$  \\
$d$           & $-0.29$ & $-0.31$     & $-0.26$ \\
    \hline
  \end{tabular}
  \caption{Mode and 68.3\% highest posterior density interval for the
    parameters under the LADE model.}
  \label{tab:HPD-LADE}
\end{table}

\begin{table}[t]
  \centering
  \begin{tabular}{lrrr}
    \hline\hline
    Parameter & Mode    & \multicolumn{2}{c}{68.3\% HPD interval} \\
              &         & $\qquad$min & max                       \\
    \hline
$\Log\,A$     & -5.67    & -5.75        & -5.54                      \\   
$\gamma_1$    & 0.90    & 0.84        & 0.94                      \\
$\gamma_2$    & 2.51    & 2.42        & 2.60                      \\
$\Log\, L_*$  & 44.05   & 43.97       & 44.12                     \\
$z_c$         & 2.10    & 2.05        & 2.19                      \\
$p_1$         & 5.08    & 4.83        & 5.48                      \\
$p_2$         & $-1.90$ & $-2.20$     & $-1.69$                   \\
$\alpha$      & 0.39    & 0.38        & 0.41                      \\
$\Log\, L_a$  & 44.70   & 44.66       & 44.74                     \\
    \hline
  \end{tabular}
  \caption{Mode and 68.3\% highest posterior density interval for the
    parameters under the LDDE model.}
  \label{tab:HPD-LDDE}
\end{table}

From the same sets of draws discussed above we produced the following
plots of the LF. In Fig.~\ref{fig:binnedLF} we plot the 68.3\% and
99.7\% (``$3\sigma$'') pointwise HPD intervals of the LF derived from
all surveys together under either the LADE or the LDDE model. A comparison with the binned estimate reveals that the LADE and LDDE
modes offer a largely overlapping description of the LF at $z\gtrsim
1$.

Some differences appear at low redshift ($z\lesssim 1$) where LADE and
LDDE appear to under-predict the LF at $L\gtrsim 10^{44}$ and
$L\gtrsim 10^{45}$ \ergs, respectively. A least-squares fit to the LF
bins at luminosities brighter than the LF knee
(Table~\ref{tab:binned-fit}) shows a progressive steepening of the
bright tail of the LF. The same effect is seen also in M15, albeit not
as clearly as here. The LADE model
(Eqs.~\ref{eq:lade1}--\ref{eq:lade4}) does not allow the double
power-law slopes to change with redshift; the LDDE model
(Eqs.~\ref{eq:ldde1}--\ref{eq:ldde3}) allows some change but which in
practice looks insufficient to follow the present data; therefore LADE
has the largest deviations. The bright tail slope is determined by the
sources with $z>1$, which are the majority at bright luminosities (see
Fig.~\ref{fig:zlum}), so the behaviour of the tail slope at lower redshifts is only
loosely constrained by our data (for this reason, LADE nonetheless
provides a better description of our data; see
Sect.~\ref{sec:ldde-vs-lade}).

\begin{table}[t]
  \centering
  \begin{tabular}{cll}
    \hline\hline
    $z$    &$\Log L_X$    &$\quad\gamma$      \\
    \hline
    $<0.5$     &$>43.5$      &$-1.3\pm 0.1$ \\
    $0.5<z<1$  &$>44$        &$-1.5\pm 0.1$ \\
    $1<z<1.5$  &$>44$        &$-1.8\pm 0.2$ \\
    $1.5<z<2$  &$>44.5$      &$-2.0$        \\
    $2<z<3$    &$>44.5$      &$-1.8$        \\
    $3<z<4$    &$>44.5$      &$-2.3$        \\
    \hline
  \end{tabular}
  \caption{Least-squares fits to the bright tail of the binned LF in
    the six redshift bins of Fig.~\ref{fig:binnedLF}. The first column
    identifies the redshift bin; the second column shows the
    luminosity range used for the fit; the third column gives the fit
    parameter $\gamma$ for the linear formula $\Log(\de \Phi/\de\Log
    L_X) = \gamma\, \Log L_X + c$. The bins with $z>1.5$ do not show an
    error for $\gamma$ because only two data points were available for
    each fit; however, an error of the same order of that of the previous
    bins may be reasonably assumed.}
  \label{tab:binned-fit}
\end{table}

Some discrepancies also appear in some redshift bins at the lowest
luminosities. The large error bars on the binned LF show that the
number of objects in these bins is limited; any discrepancy is 
contained within $2\sigma$ anyway.

The posterior densities for the double power-law parameters ($A$,
$\gamma_1$, $\gamma_2$, and $L_*$) from all surveys together under the
LADE or LDDE model are plotted as histograms in
Fig.~\ref{fig:postmodels}. The two models yield differences in $A$ and
$L_*$ of 2 and 1 order of magnitudes, respectively. This may be
explained by noting that the double power-law parameters are coupled: a
larger $L_*$ needs a smaller $A$ and a steeper slope at $L<L_*$,
which is indicated by the difference between LADE and LDDE for the left peaks
of $\gamma_1$ and $\gamma_2$.

The parameters $\gamma_1$ and $\gamma_2$ have identical, double-peaked
posteriors because they can be exchanged in Eq.~(\ref{eq:doublepow})
with no effect on the LF. The LDDE and LADE fully agree on the slope
at $L>L_*$, whose average and $1\sigma$ dispersion are $\gamma= 2.50\pm
0.13$ and $2.50\pm 0.09$ for LDDE and LADE, respectively. This value is
slightly less steep than that quoted by A10 for their colour
pre-selected sample ($\gamma_2=2.80\pm 0.12$), but it is within the
$1\sigma$ uncertainty for  the X-ray-only sample of A10 ($\gamma_2=2.36\pm
0.15$) under the LDDE model. Both U14 and M15 quote steeper slopes
(U14: $\gamma_2=2.71\pm 0.09$; M15: $\gamma_2=2.77\pm 0.12$). A less
steep slope at $L>L_*$ may result from the absorption corrections, if
a larger fraction of heavily-absorbed objects is allowed.

So far we have
commented qualitatively on the LF features; further quantitative
evaluation of the differences among the models and surveys is
presented in Sect.~\ref{sec:modelcomparison}.

\subsection{Differences among individual surveys}

The different surveys may exhibit some variance in the LF
parameters. Possible reasons include, for example, the presence of
large-scale structures (or voids) in the surveyed volume, small number
effects at the edges of the luminosity and redshift intervals,
residual effects of data reduction and source detection. In
Figs.~\ref{fig:LDDELFs} and \ref{fig:LADELFs} we plot the 68.3\% HPD
intervals under the LDDE and LADE models, respectively, for each
individual survey. The largest discrepancies appear at the
edges of the luminosity and redshift bins where differences of up to one
order of magnitude are present. The knee region is, apart from the
lowest redshift panel, the area where the different surveys agree
best. There seems to be more variance under the LADE model, where the
XMM-LSS is consistently steeper at $L\gtrsim L_*$.  

The areas where some discrepancies appear are subject to larger errors
because of the low number of objects in the relevant ranges of
luminosities and redshift: namely, the very low- and very high-luminosity bins at all redshifts.  The XMM-CDFS LF seems to be less
steep than the two others at $L\gtrsim L_*$,  which is probably because of larger
amounts of obscured objects.; this is illustrated in
Figs.~\ref{fig:postLDDE} and \ref{fig:postLADE}, where the posterior
densities for all parameters are plotted, under the LDDE and LADE
models, respectively. The XMM-CDFS also requires a lower $L_*$, by a
factor 3--10, than the two other surveys; this probably reflects the
better sampling of intrinsically fainter objects by the XMM-CDFS. The
critical redshift $z_c$ at which the rate of evolution changes is found to be
in the 1.5--2.5 interval for LDDE; it is less well constrained for
LADE. The XMM-COSMOS data do not seem to  require a decrease in the LF
after $z_c$, as hinted by the LADE $d$ and  LDDE $p_2$ parameters,
which are both  consistent with  zero.

\section{Model comparison}
\label{sec:modelcomparison}

An important reason why Bayesian inference enables a powerful model
comparison is that it naturally includes the idea of Occam's razor, which is
that among competing models predicting a similar outcome, one should
choose that with the fewest assumptions. Several metrics for model
comparison have been devised in the literature, which contain
integrals over the parameter volume (either prior, or posterior). A
model with fewer parameters than another  also has a smaller
parameter volume; and a model whose parameters are all
well-constrained occupies a smaller volume than a model with
unconstrained parameters. It is in this way that Occam's razor is
incorporated.

Bayesian evidence, also called marginal likelihood, is the
integral of the data likelihood over the prior volume. Bayesian evidence was used by
A10 to estimate that LDDE was performing slightly better than
LADE, and by A15 to reckon that LADE was not only significantly better
than LDDE, but also that LDDE15 (see Sect.~\ref{sec:ldde-ueda14}) was preferred over LADE.

Evidence is an effective metric when the priors can be easily defined,
especially in their tails \citep{trotta2008}. For LF, however, the
prior choices are still somewhat subjective. For example, the question arises as to which interval
should be permitted for $L_*$ in the case of flat priors; log-normal,
Cauchy, or exponential priors could be equally or more justified and
effective, but it is difficult to tune their parameters in an
objective manner. Therefore, in the following we prefer to focus on
metrics that are based on the data likelihood given the posterior
distribution.

We introduce the Watanabe-Akaike information criterion
\citep[WAIC;][]{watanabe2010}. It is one of a family of criteria that
estimate the predictive power of a model, i.e.\ how a model can
anticipate new data,   can be extrapolated into unobserved regions
of the luminosity-redshift space, or suggest future observations to
improve the model weaknesses.  A comprehensive review of the Bayesian
methods is in \citet[chapter 7]{BDA3} (see also \citealt{gelman2014});
an application of two of such methods to the problem of LF fitting is in
\citet{fotopoulou2016}. The underlying idea is to compute the data
likelihood under more than one model and compare these likelihoods after accounting for
the different number of parameters\footnote{The main difference
  between the WAIC and the better known, similarly named Akaike
  information criterion \citep[AIC;][]{akaike1973} used in
  \citet{fotopoulou2016}, is that the WAIC averages over the posterior
  distribution of the parameters, while AIC uses the
  maximum-likelihood estimate.}.  The ``information criterion'' part
of the name comes from the following reasoning: since every model is
only an approximation of reality, then different models can cause
different losses of information with respect to reality. Therefore, one
should choose the model that preserves most information. This is the
same concept as having more predictive power.

\smallskip

The WAIC method takes advantage of the fact
that our data are naturally partitioned with each survey representing
one partition. 
The starting point is the data
log-likelihood, averaged over the posterior distribution of the
parameters $\theta$ of model $M$. For consistency with \citet{BDA3},
we call it `log pointwise predictive density' (lppd), defined as
\begin{equation}
  \label{eq:lppd}
  \mathrm{lppd} = \ln \left( \frac{1}{S}
    \sum_{s=1}^S \prob(y|\theta^s) \right)
,\end{equation}
where each $\theta^s$ is a sample from the posterior distribution of
the parameters, $S$ is the number of posterior samples, $y$ is the
observed data, and $P(y|\theta^s)$ is the data likelihood given the
parameters $\theta^s$. We can rewrite lppd after partitioning the
data
\begin{equation}
  \label{eq:lppd2}
  \mathrm{lppd} = \sum_{i=1}^n \ln \left( 
             \frac{1}{S} \sum_{s=1}^S \prob(y_i|\theta^s)
                                          \right)
,\end{equation}
which highlights the contribution from each survey $i$ of  $n=3$
considered here, whose data are represented by $y_i$.  We obtain
$S=1000$ and use the same posterior draws from which
Figs.~\ref{fig:LDDELFs}--\ref{fig:postLADE} are plotted.

\begin{table}
  \centering
  \begin{tabular}{lr}
    \hline\hline
    Model   &$\Delta$ WAIC  \\
    \hline
    LADE    &0     \\
    LDDE    &27    \\
    LDDE15  &105   \\

    \hline
  \end{tabular}
  \caption{Differences between the values of the Watanabe-Akaike
    information criterion, for all surveys together under the LADE,
    LDDE, and LDDE15 models. The zero is set to the model with the
    lower WAIC. The lower the WAIC, the larger the
    predictive power of a model. A number of 1000 draws from the posterior
    distribution were used. }
  \label{tab:WAIC}
\end{table}

The WAIC operates by adding a correction $p_\mathrm{WAIC}$ to lppd to further
adjust for the number of parameters. The correction is the variance of
lppd among the different surveys as follows:
\begin{equation}
  \label{eq:pwaic}
  p_\mathrm{WAIC} = \sum_{i=1}^n \frac{1}{S-1} \sum_{s=1}^S 
      \left(
        \ln \prob (y_i|\theta^s) - <\ln \prob(y_i|\theta^s)> 
      \right)^2
,\end{equation}
where the angular brackets $<>$ indicate that the average over $s$ should be taken.

The WAIC is finally defined as
\begin{equation}
  \label{eq:WAIC}
  \mathrm{WAIC} = -2 \left( \mathrm{lppd} - p_\mathrm{WAIC} \right)
  \quad .
\end{equation}

Absolute WAIC values are not relevant since they are
dominated by the sample size; only the differences
 carry statistical meaning.
The $\Delta$WAIC values for the LDDE and LADE models are shown in
Table~\ref{tab:WAIC}. The lower the WAIC, the more predictive power
 the model has. A difference of 27 can be
observed between LDDE and LADE, leading to a preference for
LADE. Table~\ref{tab:WAIC} also shows LDDE15, which is discussed in
the next Section.

\section{On a proposed extension to LDDE}
\label{sec:ldde-ueda14}

In U14, we present an extension to the LDDE model. The main
difference is the inclusion of a second break at a
redshift $z>z_0$. This adds six more parameters, making a total of 15
parameters for the double power law plus evolution. This model, which
we call here LDDE15, takes the following functional form%
\footnote{A similar extension is considered in M15, where  $p_2$ also has a
dependence on $L$ (like Eq.~\ref{eq:ldde15_2}, but with its own
exponent $\beta_2$), but where $\alpha_2$ and $L_{\alpha2}$ are
missing. In M15, as in U14, some parameters are fixed in the fit.} (compare with
Eqs.~\ref{eq:ldde2}, \ref{eq:ldde3}):
\begin{equation}
\label{eq:ldde15_1}
\begin{array}{l}
\mathrm{LDDE15}(L,z) = \\
\mbox{\hspace{1ex}} = \left\{ \!\! \begin{array}{ll}
  (1+z)^{p_1}                                       &z \le z_1(L)  \\
  (1+z_1)^{p_1}\!\!
     \left(\frac{1+z}{1+z_1}\right)^{p_2}\quad     &z_1(L) < z \le
     z_1(L)  \\
  (1+z_1)^{p_1}\!\!
     \left(\frac{1+z}{1+z_1}\right)^{p_2}\!\!
     \left(\frac{1+z}{1+z_2}\right)^{p_3}\quad     &z > z_2(L)
\end{array}  \right.
\end{array}
\end{equation}
with
\begin{equation}
\label{eq:ldde15_2}
p_1(L) = p_1^* + \beta_1 (\log\ L - \log\ L_p)
\end{equation}
\begin{equation}
\label{eq:ldde15_3}
z_1(L) = \left\{ \begin{array}{ll}
    z_{c1}                                                & L \ge L_{\alpha1} \\
    z_{c1} \left( \frac{L}{L_{\alpha1}} \right)^{\alpha1} \qquad  & L < L_{\alpha1} \quad ,
  \end{array} \right.
\end{equation}
\begin{equation}
\label{eq:ldde15_4}
z_2(L) = \left\{ \begin{array}{ll}
    z_{c2}                                                & L \ge L_{\alpha2} \\
    z_{c2} \left( \frac{L}{L_{\alpha2}} \right)^{\alpha2} \qquad  & L < L_{\alpha2} \quad .
  \end{array} \right.
\end{equation}

In U14, $p_3$, $z_{c_2}$, $\alpha_2$, and $L_{\alpha_2}$ were fixed at
values that make the LF decline faster beyond $z\sim 3$, reproducing
the results by \citep{fiore2012}.  A15 considered this
model and used Bayesian methods to check whether all parameters could
be constrained by the data, finding reasonably tight dispersions
around the means for most of the parameters. Parameters $L_p$ and $p_3$ seem to have looser
constraints.

We have also used Bayesian inference with the LDDE15 model, using the
same priors of A15. These priors are slightly more
informative than what used in Sect.~\ref{sec:bayes}; the main
difference is that $\gamma_1$ and $\gamma_2$ are no longer bimodal
and that the allowed intervals for $p_1$, $p_2$, and $p_3$, and for
$z_{c1}$ and $z_{c2}$, are not overlapping.

\begin{table}[t]
  \centering
  \begin{tabular}{lrrr}
    \hline\hline
    Parameter         & Mode      & \multicolumn{2}{c}{68.3\% HPD interval} \\
                      &           & $\qquad$min & max                       \\
    \hline
$\Log\,A$             & -5.37      & -5.43        & -5.26                   \\ 
$\gamma_1$            & 0.89      & 0.84        & 0.93                      \\ 
$\gamma_2$            & 2.90      & 2.78        & 3.02                      \\ 
$\Log\, L_*$          & 44.08     & 44.03       & 44.12                     \\ 
$z_{c1}$              & 2.21      & 2.15        & 2.27                      \\ 
$p_1^*$               & 3.57      & 2.61        & 5.56                      \\ 
$\beta_1$             & 1.48      & 1.17        & 1.73                      \\ 
$L_p$                 & \multicolumn{3}{c}{unconstrained}                   \\
$p_2$                 & \multicolumn{3}{c}{unconstrained}                   \\
$\alpha_1$            & \multicolumn{3}{c}{unconstrained}                   \\
$\Log\, L_{\alpha 1}$ & \multicolumn{3}{c}{unconstrained}                   \\
$z_{c2}$              & 2.77      & 2.61        & 2.95                      \\
$p_3$                 & $-5.00^*$ & $-5.13$     & $-5.00$                   \\
$\alpha_2$            & 0.30      & 0.28        & 0.31                      \\
$\Log\, L_{\alpha 2}$ & 44.66     & 44.62       & 44.71                     \\
    \hline
  \end{tabular}
  \caption{Mode and 68.3\% highest posterior density interval for the
    parameters under the LDDE15 model.
    \newline
    $^*$ The mode for $p_3$ corresponds to the upper bound of the
    allowed range.
  }
  \label{tab:HPD-LDDE15}
\end{table}

\begin{figure*}
  \centering
  \includegraphics[height=.9\textwidth,angle=-90]{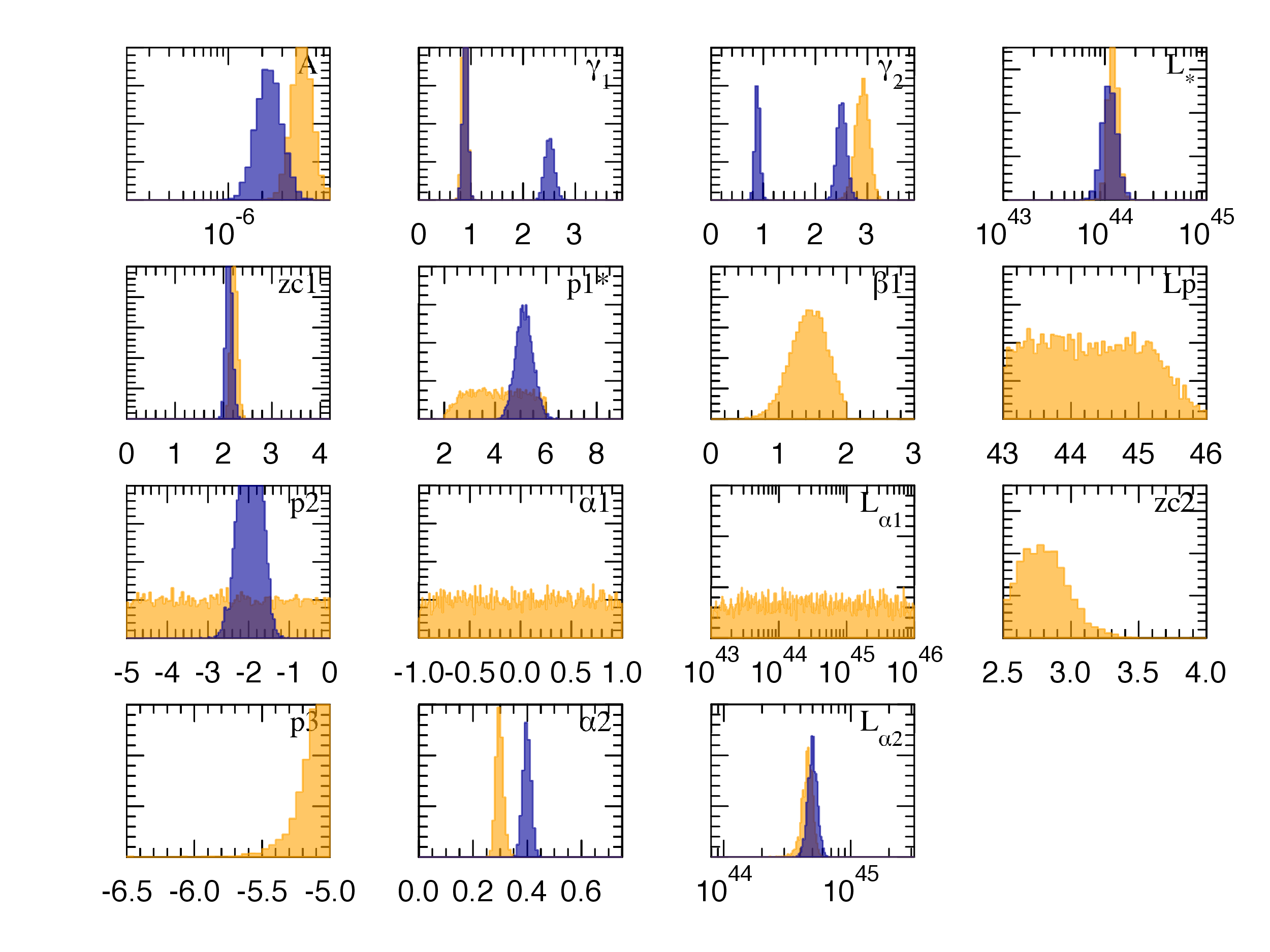}
  \caption{ Posterior probability densities for double power-law and
    evolution parameters for all surveys together under the LDDE
    (blue) and  LDDE15 (orange) model.}
  \label{fig:postLDDE15}
\end{figure*}

The mode and HPD intervals of our posterior densities for all
parameters are shown in Table~\ref{tab:HPD-LDDE15}. In
Fig.~\ref{fig:postLDDE15} we plot the histograms of the posterior
densities; for the parameters  in common with LDDE we also
plot the LDDE histograms for comparison. Most of the parameters in
common have similar (within a few full width half maximums) posterior
densities. The role of LDDE's $\alpha$ and $L_\alpha$ (shaping the
luminosity-dependent decrease of the LF at high redshift) is taken by
$\alpha_2$ and $L_{\alpha2}$ in LDDE15, which assume similar
values. We find $p_1^* > p_1$, and $0\lesssim \beta_1 \lesssim 1$, but
$L_p$ cannot be constrained.  The second critical redshift ($z_{c2}$)
can be constrained in the 2.6--3.7 interval (to be compared with
$2.1\lesssim z_{c1}\lesssim 2.2$). The other parameters ($L_p$, $p_2$,
$p_3$, $\alpha_1$, and $L_{\alpha1}$) could either not be constrained
at all, or only mildly constrained ($p_3$, whose posterior is very
skewed towards the boundary).

In summary, of 15 parameters in the LDDE15 model, only ten could be
fully constrained, and another is skewed towards its boundary.  Also,
the WAIC for the LDDE15 model is larger
(Table~\ref{tab:WAIC}). This is due to the presence in LDDE15 of
parameters with unconstrained posterior densities, which enlarge the
parameter volume.  Our interpretation is that the data discussed in
this paper do not support this particular extension of the LDDE model.

\section{Discussion}
\label{sec:discussion}

\subsection{LDDE vs.\ LADE}
\label{sec:ldde-vs-lade}

Our findings appear to be at odds with  A15 found (U14 did not
compare LDDE with LDDE15).  It is unclear what is causing this
difference. A15  probably has a somewhat better sampling of the LF in
the $3<z<4$ redshift interval (from their Fig.~3 we count $\sim 140$
objects,  compared with 97 in our samples), but the difference in
the number of objects looks too small to justify the different
results. Photometric redshifts were computed by A15 using templates
and codes that are different from what we used. Most
notably, the photometric redshifts used here are tuned for X-ray
sources in that\ they include a bias towards AGN rather than galaxies;
while A15 uses a more general set of templates that may give less
accurate redshifts and a larger outlier rate (see their
Sect.~2.6). The possibility of cosmic variance explaining the
different results seems unlikely, given that some fields (CDFS and
COSMOS) are common between A15 and this work.

However,  A15 also attempt to build a nested model
for LF evolution (flexible double power law). The underlying idea of this model is
closer to LADE than LDDE: a polynomial characterisation on $z$ is put
on each of the four double power-law parameters. This allows these authors to investigate
up to what orders the polynomial coefficients can be constrained. They
find a maximum of ten parameters (ten is the sum of the orders of all four
polynomials). Constraining at most ten parameters looks closer to our
results.

It is possible that future larger surveys, most notably the XXL, or an
increase in the number of spectroscopic redshifts might yield
different results. In the meantime, non-parametric methods, such as our
formulation in Sect.~\ref{sec:binnedLF} or the interesting Bayesian
adaptation by \cite{buchner2015}, will continue to play an important role in
understanding how future models should be shaped.

\begin{figure*}
  \centering
  \includegraphics[height=.9\textwidth,angle=-90]{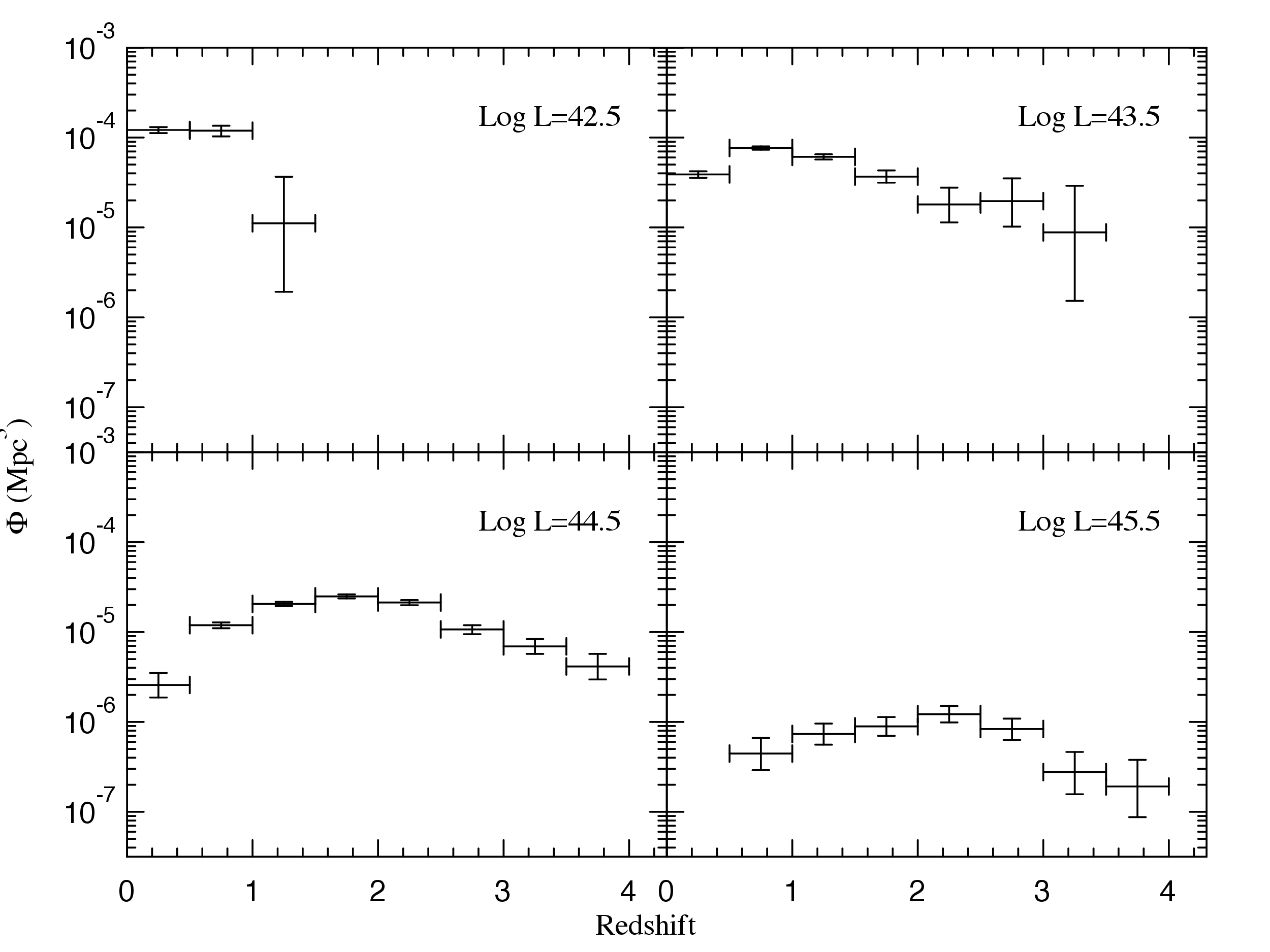}
  \caption{Redshift distribution for all surveys combined, in four
    different bins of luminosity (the average $\Log L$ of the bin shown in each panel). The distribution peaks at increasing
    redshift for higher luminosities, illustrating the downsizing of
    black hole accretion during the lifetime of the universe.  }
  \label{fig:zdistr}
\end{figure*}

\subsection{Redshift distribution}
\label{sec:zdistr}

The LF can be plotted in terms of redshift to show the
luminosity-dependent redshift distribution (RD).  In
Fig.~\ref{fig:zdistr}, we present  the binned RD from all surveys combined in four
bins of luminosity ($42\le \Log L<43$, $43\le \Log L<44$,
$44\le \Log L<45$, and $45\le \Log L<46$). The peak of the RD depends on
the luminosity bin; at increasing luminosities, the peak is found at higher
redshift. Our data clearly show the downsizing of BH growth,
i.e.\ the idea that accretion was happening on more massive scales at
larger redshift \citep{cowie1996,ueda03,hasinger2005}. For the highest
luminosity bin that we considered (average $\Log L=45.5$), the peak of
accretion happens between $z\sim 2.0$ and 2.5. For the lowest
luminosity bin (average $\Log L=42.5$), an upper limit to the peak
could be put at $z\lesssim 0.5$. The data are overall consistent,
within errors, with other determinations (e.g.\ U14, M15).

\section{Conclusions}
\label{sec:conclusion}

We have presented the luminosity function (LF) estimated from the XMM-LSS,
XMM-COSMOS, and XMM-CDFS surveys, and from their combination. A total
of 2887 AGN is used to build the LF in the luminosity interval
$10^{42}$--$10^{46}$ \ergs and in the redshift interval 0.001-4.

We presented a method to  account for absorption statistically, based on the
probability distribution of the absorber column density conditioned
on the soft/hard flux ratio. We apply this method
both to non-parametric estimates, modifying the \citep{pageca00}
method, and to parametric estimates, introducing the corrections in
the likelihood formula.

We  presented both non-parametric and parametric estimates of the
LF. The parametric form is a double power law with either LADE or LDDE
evolution. Bayesian inference methods allow us to obtain a full and
reliable characterisation of the allowed parameter space. The full
posterior probability density for both the LF and the LF parameters is
shown for both the LADE and LDDE models. The results are consistent,
within errors, with previous literature. A comparison between the
non-parametric and parametric estimates
reveals that the LADE and LDDE modes offer a largely
overlapping description of the LF at $z\gtrsim 1$. Some differences
appear at low redshift ($z\lesssim 0.5$), where LADE appears to
under-predict the LF at $L\gtrsim 10^{44}$ \ergs. Difference also exist in each
redshift bin at the lowest luminosities, however, where  the number of
objects in the lower luminosity bin is small so uncertainties are
large.

We  introduced the use of the fully Bayesian, Watanabe-Akaike
information criterion (WAIC) to compare the predictive power of
different models. The LADE model is found to have more predictive
power than the LDDE model.
The difference in WAIC values between the two models can be
interpreted as a measure of how better one model can describe future and/or
out-of-sample data.
We have investigated the 15-parameter extended LDDE
model (LDDE15), finding that our data do not support this
extension. Among the possible explanation for this discrepancy between
our results and \cite{ueda2014}, we mention a different approach in
computing photometric redshift, and the different sample sizes in the
$3<z<4$ redshift range.
The binned LF, plotted as a redshift distribution, clearly illustrates
the downsizing of black hole accretion, which is in agreement with previous
studies.

\begin{acknowledgements}
  We thank an anonymous referee whose comments have contributed to
  improving the presentation of this paper.
 PR thanks Zhang Xiaoxia for reporting typos.
PR acknowledges a grant from
the Greek General Secretariat of Research and Technology in the
framework of the programme Support of Postdoctoral Researchers.
FJC acknowledges
support by the Spanish ministry of Economy and Competitiveness through
the grant AYA2012-31447.

\end{acknowledgements}

\bibliographystyle{aa}
\bibliography{../fullbiblio}

\begin{thebibliography}{86}
\expandafter\ifx\csname natexlab\endcsname\relax\def\natexlab#1{#1}\fi

\bibitem[{{Aird} {et~al.}(2015){Aird}, {Coil}, {Georgakakis}, {Nandra},
  {Barro}, \& {P{\'e}rez-Gonz{\'a}lez}}]{aird2015}
{Aird}, J., {Coil}, A.~L., {Georgakakis}, A., {et~al.} 2015, \mnras, 451, 1892

\bibitem[{{Aird} {et~al.}(2010){Aird}, {Nandra}, {Laird}, {Georgakakis},
  {Ashby}, {Barmby}, {Coil}, {Huang}, {Koekemoer}, {Steidel}, \&
  {Willmer}}]{aird2010}
{Aird}, J., {Nandra}, K., {Laird}, E.~S., {et~al.} 2010, \mnras, 401, 2531

\bibitem[{Akaike(1973)}]{akaike1973}
Akaike, H. 1973, in 2nd International Symposium on Information Theory,
  Academiai Kiado, reprinted in Selected Papers of Hirotugu Akaike (Springer
  1998), 199--213

\bibitem[{{Akylas} {et~al.}(2012){Akylas}, {Georgakakis}, {Georgantopoulos},
  {Brightman}, \& {Nandra}}]{akylas2012}
{Akylas}, A., {Georgakakis}, A., {Georgantopoulos}, I., {Brightman}, M., \&
  {Nandra}, K. 2012, \aap, 546, A98

\bibitem[{{Alexander} \& {Hickox}(2012)}]{alexander2012}
{Alexander}, D.~M. \& {Hickox}, R.~C. 2012, \nar, 56, 93

\bibitem[{{Andreon}(2012)}]{andreon2011}
{Andreon}, S. 2012, in Astrostatistical Challenges for the New Astronomy, ed.
  J.~{Hilbe} (Springer), 41--62, also available as arXiv:1112.3652

\bibitem[{Andreon \& Hurn(2013)}]{andreon2013}
Andreon, S. \& Hurn, M. 2013, Statistical Analysis and Data Mining, 6, 15, also
  available as arXiv:1210.6232

\bibitem[{{Arnouts} {et~al.}(1999){Arnouts}, {Cristiani}, {Moscardini},
  {Matarrese}, {Lucchin}, {Fontana}, \& {Giallongo}}]{arnouts1999-lephare}
{Arnouts}, S., {Cristiani}, S., {Moscardini}, L., {et~al.} 1999, \mnras, 310,
  540

\bibitem[{{Ballantyne} {et~al.}(2006){Ballantyne}, {Everett}, \&
  {Murray}}]{ballantyne2006}
{Ballantyne}, D.~R., {Everett}, J.~E., \& {Murray}, N. 2006, \apj, 639, 740

\bibitem[{{Barger} {et~al.}(2005){Barger}, {Cowie}, {Mushotzky}, {Yang},
  {Wang}, {Steffen}, \& {Capak}}]{barger05}
{Barger}, A.~J., {Cowie}, L.~L., {Mushotzky}, R.~F., {et~al.} 2005, \aj, 129,
  578

\bibitem[{{Boyle} {et~al.}(1993){Boyle}, {Griffiths}, {Shanks}, {Stewart}, \&
  {Georgantopoulos}}]{boyle1993}
{Boyle}, B.~J., {Griffiths}, R.~E., {Shanks}, T., {Stewart}, G.~C., \&
  {Georgantopoulos}, I. 1993, \mnras, 260, 49

\bibitem[{{Boyle} {et~al.}(1994){Boyle}, {Shanks}, {Georgantopoulos},
  {Stewart}, \& {Griffiths}}]{boyle1994}
{Boyle}, B.~J., {Shanks}, T., {Georgantopoulos}, I., {Stewart}, G.~C., \&
  {Griffiths}, R.~E. 1994, \mnras, 271, 639

\bibitem[{{Braccesi} {et~al.}(1980){Braccesi}, {Zitelli}, {Bonoli}, \&
  {Formiggini}}]{braccesi1980}
{Braccesi}, A., {Zitelli}, V., {Bonoli}, F., \& {Formiggini}, L. 1980, \aap,
  85, 80

\bibitem[{{Brightman} \& {Ueda}(2012)}]{brightman2012}
{Brightman}, M. \& {Ueda}, Y. 2012, \mnras, 423, 702

\bibitem[{{Buchner} {et~al.}(2015){Buchner}, {Georgakakis}, {Nandra},
  {Brightman}, {Menzel}, {Liu}, {Hsu}, {Salvato}, {Rangel}, {Aird}, {Merloni},
  \& {Ross}}]{buchner2015}
{Buchner}, J., {Georgakakis}, A., {Nandra}, K., {et~al.} 2015, \apj, 802, 89

\bibitem[{{Burlon} {et~al.}(2011){Burlon}, {Ajello}, {Greiner}, {Comastri},
  {Merloni}, \& {Gehrels}}]{burlon2011}
{Burlon}, D., {Ajello}, M., {Greiner}, J., {et~al.} 2011, \apj, 728, 58

\bibitem[{{Cappelluti} {et~al.}(2009){Cappelluti}, {Brusa}, {Hasinger},
  {Comastri}, {Zamorani}, {Finoguenov}, {Gilli}, {Puccetti}, {Miyaji},
  {Salvato}, {Vignali}, {Aldcroft}, {B{\"o}hringer}, {Brunner}, {Civano},
  {Elvis}, {Fiore}, {Fruscione}, {Griffiths}, {Guzzo}, {Iovino}, {Koekemoer},
  {Mainieri}, {Scoville}, {Shopbell}, {Silverman}, \& {Urry}}]{xmm-cosmos}
{Cappelluti}, N., {Brusa}, M., {Hasinger}, G., {et~al.} 2009, \aap, 497, 635

\bibitem[{{Chiappetti} {et~al.}(2013){Chiappetti}, {Clerc}, {Pacaud}, {Pierre},
  {Gu{\'e}guen}, {Paioro}, {Polletta}, {Melnyk}, {Elyiv}, {Surdej}, \&
  {Faccioli}}]{chiappetti2013}
{Chiappetti}, L., {Clerc}, N., {Pacaud}, F., {et~al.} 2013, \mnras, 429, 1652

\bibitem[{{Comastri} {et~al.}(1995){Comastri}, {Setti}, {Zamorani}, \&
  {Hasinger}}]{comastri95}
{Comastri}, A., {Setti}, G., {Zamorani}, G., \& {Hasinger}, G. 1995, \aap, 296,
  1

\bibitem[{{Cowie} {et~al.}(1996){Cowie}, {Songaila}, {Hu}, \&
  {Cohen}}]{cowie1996}
{Cowie}, L.~L., {Songaila}, A., {Hu}, E.~M., \& {Cohen}, J.~G. 1996, \aj, 112,
  839

\bibitem[{{Daddi} {et~al.}(2007){Daddi}, {Alexander}, {Dickinson}, {Gilli},
  {Renzini}, {Elbaz}, {Cimatti}, {Chary}, {Frayer}, {Bauer}, {Brandt},
  {Giavalisco}, {Grogin}, {Huynh}, {Kurk}, {Mignoli}, {Morrison}, {Pope}, \&
  {Ravindranath}}]{daddi2007}
{Daddi}, E., {Alexander}, D.~M., {Dickinson}, M., {et~al.} 2007, \apj, 670, 173

\bibitem[{{Ebrero} {et~al.}(2009){Ebrero}, {Carrera}, {Page}, {Silverman},
  {Barcons}, {Ceballos}, {Corral}, {Della Ceca}, \& {Watson}}]{ebrero2009}
{Ebrero}, J., {Carrera}, F.~J., {Page}, M.~J., {et~al.} 2009, \aap, 493, 55

\bibitem[{{Feroz} \& {Hobson}(2008)}]{feroz2008multinest}
{Feroz}, F. \& {Hobson}, M.~P. 2008, \mnras, 384, 449

\bibitem[{{Feroz} {et~al.}(2009){Feroz}, {Hobson}, \&
  {Bridges}}]{feroz2009multinest}
{Feroz}, F., {Hobson}, M.~P., \& {Bridges}, M. 2009, \mnras, 398, 1601

\bibitem[{{Feroz} {et~al.}(2013){Feroz}, {Hobson}, {Cameron}, \&
  {Pettitt}}]{feroz2013multinest}
{Feroz}, F., {Hobson}, M.~P., {Cameron}, E., \& {Pettitt}, A.~N. 2013, ArXiv
  e-prints

\bibitem[{{Ferrarese} \& {Merritt}(2000)}]{ferrares2000}
{Ferrarese}, L. \& {Merritt}, D. 2000, \apjl, 539, L9

\bibitem[{{Fiore} {et~al.}(2012){Fiore}, {Puccetti}, {Grazian}, {Menci},
  {Shankar}, {Santini}, {Piconcelli}, {Koekemoer}, {Fontana}, {Boutsia},
  {Castellano}, {Lamastra}, {Malacaria}, {Feruglio}, {Mathur}, {Miller}, \&
  {Pannella}}]{fiore2012}
{Fiore}, F., {Puccetti}, S., {Grazian}, A., {et~al.} 2012, \aap, 537, A16

\bibitem[{{Fontanot} {et~al.}(2009){Fontanot}, {De Lucia}, {Monaco},
  {Somerville}, \& {Santini}}]{fontanot2009}
{Fontanot}, F., {De Lucia}, G., {Monaco}, P., {Somerville}, R.~S., \&
  {Santini}, P. 2009, \mnras, 397, 1776

\bibitem[{{Fotopoulou} {et~al.}(2016){Fotopoulou}, {Buchner},
  {Georgantopoulos}, {Hasinger}, {Salvato}, {Georgakakis}, {Cappelluti},
  {Ranalli}, {Hsu}, {Brusa}, {Comastri}, {Miyaji}, {Nandra}, {Aird}, \&
  {Paltani}}]{fotopoulou2016}
{Fotopoulou}, S., {Buchner}, J., {Georgantopoulos}, I., {et~al.} 2016, \aap,
  587, A142

\bibitem[{{Gebhardt} {et~al.}(2000){Gebhardt}, {Bender}, {Bower}, {Dressler},
  {Faber}, {Filippenko}, {Green}, {Grillmair}, {Ho}, {Kormendy}, {Lauer},
  {Magorrian}, {Pinkney}, {Richstone}, \& {Tremaine}}]{gebhardt2000}
{Gebhardt}, K., {Bender}, R., {Bower}, G., {et~al.} 2000, \apjl, 539, L13

\bibitem[{{Gehrels}(1986)}]{gehrels86}
{Gehrels}, N. 1986, \apj, 303, 336

\bibitem[{Gelman {et~al.}(2013)Gelman, Carlin, Stern, Dunson, Vehtari, \&
  Rubin}]{BDA3}
Gelman, A., Carlin, J., Stern, H., {et~al.} 2013, Bayesian Data Analysis, Third
  Edition, Chapman \& Hall/CRC Texts in Statistical Science (Taylor \& Francis)

\bibitem[{Gelman {et~al.}(2014)Gelman, Hwang, \& Vehtari}]{gelman2014}
Gelman, A., Hwang, J., \& Vehtari, A. 2014, Statistics and Computing, 24, 997

\bibitem[{{Gilli} {et~al.}(2007){Gilli}, {Comastri}, \& {Hasinger}}]{gilli2007}
{Gilli}, R., {Comastri}, A., \& {Hasinger}, G. 2007, \aap, 463, 79

\bibitem[{Gregory(2005)}]{gregory2005bayesian}
Gregory, P. 2005, Bayesian Logical Data Analysis for the Physical Sciences: A
  Comparative Approach with Mathematica{\textregistered} Support (Cambridge
  University Press)

\bibitem[{{G{\"u}ltekin} {et~al.}(2009){G{\"u}ltekin}, {Richstone}, {Gebhardt},
  {Lauer}, {Tremaine}, {Aller}, {Bender}, {Dressler}, {Faber}, {Filippenko},
  {Green}, {Ho}, {Kormendy}, {Magorrian}, {Pinkney}, \&
  {Siopis}}]{guetelkin2009}
{G{\"u}ltekin}, K., {Richstone}, D.~O., {Gebhardt}, K., {et~al.} 2009, \apj,
  698, 198

\bibitem[{{H{\"a}ring} \& {Rix}(2004)}]{haering2004}
{H{\"a}ring}, N. \& {Rix}, H.-W. 2004, \apjl, 604, L89

\bibitem[{{Hasinger}(2008)}]{hasinger2008}
{Hasinger}, G. 2008, \aap, 490, 905

\bibitem[{{Hasinger} {et~al.}(2005){Hasinger}, {Miyaji}, \&
  {Schmidt}}]{hasinger2005}
{Hasinger}, G., {Miyaji}, T., \& {Schmidt}, M. 2005, \aap, 441, 417

\bibitem[{{Hiroi} {et~al.}(2012){Hiroi}, {Ueda}, {Akiyama}, \&
  {Watson}}]{hiroi2012}
{Hiroi}, K., {Ueda}, Y., {Akiyama}, M., \& {Watson}, M.~G. 2012, \apj, 758, 49

\bibitem[{{Hopkins} {et~al.}(2006){Hopkins}, {Hernquist}, {Cox}, {Di Matteo},
  {Robertson}, \& {Springel}}]{hopkins2006cxb}
{Hopkins}, P.~F., {Hernquist}, L., {Cox}, T.~J., {et~al.} 2006, \apjs, 163, 1

\bibitem[{{Hopkins} {et~al.}(2007){Hopkins}, {Hernquist}, {Cox}, {Robertson},
  \& {Krause}}]{hopkins2007}
{Hopkins}, P.~F., {Hernquist}, L., {Cox}, T.~J., {Robertson}, B., \& {Krause},
  E. 2007, \apj, 669, 67

\bibitem[{{Hsu} {et~al.}(2014){Hsu}, {Salvato}, {Nandra}, {Brusa}, {Bender},
  {Buchner}, {Donley}, {Kocevski}, {Guo}, {Hathi}, {Rangel}, {Willner},
  {Brightman}, {Georgakakis}, {Budav{\'a}ri}, {Szalay}, {Ashby}, {Barro},
  {Dahlen}, {Faber}, {Ferguson}, {Galametz}, {Grazian}, {Grogin}, {Huang},
  {Koekemoer}, {Lucas}, {McGrath}, {Mobasher}, {Peth}, {Rosario}, \&
  {Trump}}]{hsu2014}
{Hsu}, L.-T., {Salvato}, M., {Nandra}, K., {et~al.} 2014, \apj, 796, 60

\bibitem[{{Ilbert} {et~al.}(2006){Ilbert}, {Arnouts}, {McCracken},
  {Bolzonella}, {Bertin}, {Le F{\`e}vre}, {Mellier}, {Zamorani}, {Pell{\`o}},
  {Iovino}, {Tresse}, {Le Brun}, {Bottini}, {Garilli}, {Maccagni}, {Picat},
  {Scaramella}, {Scodeggio}, {Vettolani}, {Zanichelli}, {Adami}, {Bardelli},
  {Cappi}, {Charlot}, {Ciliegi}, {Contini}, {Cucciati}, {Foucaud}, {Franzetti},
  {Gavignaud}, {Guzzo}, {Marano}, {Marinoni}, {Mazure}, {Meneux}, {Merighi},
  {Paltani}, {Pollo}, {Pozzetti}, {Radovich}, {Zucca}, {Bondi}, {Bongiorno},
  {Busarello}, {de La Torre}, {Gregorini}, {Lamareille}, {Mathez}, {Merluzzi},
  {Ripepi}, {Rizzo}, \& {Vergani}}]{ilbert2006-lephare}
{Ilbert}, O., {Arnouts}, S., {McCracken}, H.~J., {et~al.} 2006, \aap, 457, 841

\bibitem[{{Iwasawa} {et~al.}(2012){Iwasawa}, {Gilli}, {Vignali}, {Comastri},
  {Brandt}, {Ranalli}, {Vito}, {Cappelluti}, {Carrera}, {Falocco},
  {Georgantopoulos}, {Mainieri}, \& {Paolillo}}]{iwasawa2012}
{Iwasawa}, K., {Gilli}, R., {Vignali}, C., {et~al.} 2012, \aap, 546, A84

\bibitem[{{Jones} {et~al.}(1997){Jones}, {McHardy}, {Merrifield}, {Mason},
  {Smith}, {Abraham}, {Branduardi-Raymont}, {Newsam}, {Dalton},
  {Rowan-Robinson}, \& {Luppino}}]{jones1997}
{Jones}, L.~R., {McHardy}, I.~M., {Merrifield}, M.~R., {et~al.} 1997, \mnras,
  285, 547

\bibitem[{{Kodama} {et~al.}(2004){Kodama}, {Yamada}, {Akiyama}, {Aoki}, {Doi},
  {Furusawa}, {Fuse}, {Imanishi}, {Ishida}, {Iye}, {Kajisawa}, {Karoji},
  {Kobayashi}, {Komiyama}, {Kosugi}, {Maeda}, {Miyazaki}, {Mizumoto},
  {Morokuma}, {Nakata}, {Noumaru}, {Ogasawara}, {Ouchi}, {Sasaki}, {Sekiguchi},
  {Shimasaku}, {Simpson}, {Takata}, {Tanaka}, {Ueda}, {Yasuda}, \&
  {Yoshida}}]{kodama2004}
{Kodama}, T., {Yamada}, T., {Akiyama}, M., {et~al.} 2004, \mnras, 350, 1005

\bibitem[{{Kormendy} \& {Bender}(2009)}]{kormendy2009}
{Kormendy}, J. \& {Bender}, R. 2009, \apjl, 691, L142

\bibitem[{{La Franca} \& {Cristiani}(1997)}]{lafranca1997}
{La Franca}, F. \& {Cristiani}, S. 1997, \aj, 113, 1517

\bibitem[{{La Franca} {et~al.}(2005){La Franca}, {Fiore}, {Comastri}, {Perola},
  {Sacchi}, {Brusa}, {Cocchia}, {Feruglio}, {Matt}, {Vignali}, {Carangelo},
  {Ciliegi}, {Lamastra}, {Maiolino}, {Mignoli}, {Molendi}, \&
  {Puccetti}}]{lafranca2005}
{La Franca}, F., {Fiore}, F., {Comastri}, A., {et~al.} 2005, \apj, 635, 864

\bibitem[{{Loredo}(2004)}]{loredo2004}
{Loredo}, T.~J. 2004, in American Institute of Physics Conference Series, Vol.
  735, American Institute of Physics Conference Series, ed. R.~{Fischer},
  R.~{Preuss}, \& U.~V. {Toussaint}, 195--206

\bibitem[{{Maccacaro} {et~al.}(1991){Maccacaro}, {della Ceca}, {Gioia},
  {Morris}, {Stocke}, \& {Wolter}}]{maccacaro1991}
{Maccacaro}, T., {della Ceca}, R., {Gioia}, I.~M., {et~al.} 1991, \apj, 374,
  117

\bibitem[{{Maccacaro} {et~al.}(1983){Maccacaro}, {Gioia}, {Avni}, {Giommi},
  {Griffiths}, {Liebert}, {Stocke}, \& {Danziger}}]{maccacaro1983}
{Maccacaro}, T., {Gioia}, I.~M., {Avni}, Y., {et~al.} 1983, \apjl, 266, L73

\bibitem[{{Maccacaro} {et~al.}(1984){Maccacaro}, {Gioia}, \&
  {Stocke}}]{maccacaro1984}
{Maccacaro}, T., {Gioia}, I.~M., \& {Stocke}, J.~T. 1984, \apj, 283, 486

\bibitem[{{Magorrian} {et~al.}(1998){Magorrian}, {Tremaine}, {Richstone},
  {Bender}, {Bower}, {Dressler}, {Faber}, {Gebhardt}, {Green}, {Grillmair},
  {Kormendy}, \& {Lauer}}]{magorrian1998}
{Magorrian}, J., {Tremaine}, S., {Richstone}, D., {et~al.} 1998, \aj, 115, 2285

\bibitem[{{Malizia} {et~al.}(2009){Malizia}, {Stephen}, {Bassani}, {Bird},
  {Panessa}, \& {Ubertini}}]{malizia2009}
{Malizia}, A., {Stephen}, J.~B., {Bassani}, L., {et~al.} 2009, \mnras, 399, 944

\bibitem[{{Marconi} \& {Hunt}(2003)}]{marconi2003}
{Marconi}, A. \& {Hunt}, L.~K. 2003, \apjl, 589, L21

\bibitem[{{Marconi} {et~al.}(2004){Marconi}, {Risaliti}, {Gilli}, {Hunt},
  {Maiolino}, \& {Salvati}}]{marconi2004}
{Marconi}, A., {Risaliti}, G., {Gilli}, R., {et~al.} 2004, \mnras, 351, 169

\bibitem[{{Marshall} {et~al.}(1983){Marshall}, {Tananbaum}, {Avni}, \&
  {Zamorani}}]{marshall1983}
{Marshall}, H.~L., {Tananbaum}, H., {Avni}, Y., \& {Zamorani}, G. 1983, \apj,
  269, 35

\bibitem[{{Mathez}(1978)}]{mathez1978-ple}
{Mathez}, G. 1978, \aap, 68, 17

\bibitem[{{Melnyk} {et~al.}(2013){Melnyk}, {Plionis}, {Elyiv}, {Salvato},
  {Chiappetti}, {Clerc}, {Gandhi}, {Pierre}, {Sadibekova},
  {Pospieszalska-Surdej}, \& {Surdej}}]{melnyk2013}
{Melnyk}, O., {Plionis}, M., {Elyiv}, A., {et~al.} 2013, \aap, 557, A81

\bibitem[{{Miyaji} {et~al.}(2015){Miyaji}, {Hasinger}, {Salvato}, {Brusa},
  {Cappelluti}, {Civano}, {Puccetti}, {Elvis}, {Brunner}, {Fotopoulou}, {Ueda},
  {Griffiths}, {Koekemoer}, {Akiyama}, {Comastri}, {Gilli}, {Lanzuisi},
  {Merloni}, \& {Vignali}}]{miyaji2015}
{Miyaji}, T., {Hasinger}, G., {Salvato}, M., {et~al.} 2015, \apj, 804, 104

\bibitem[{{Miyaji} {et~al.}(2001){Miyaji}, {Hasinger}, \&
  {Schmidt}}]{miyaji2001}
{Miyaji}, T., {Hasinger}, G., \& {Schmidt}, M. 2001, \aap, 369, 49

\bibitem[{{Page} \& {Carrera}(2000)}]{pageca00}
{Page}, M.~J. \& {Carrera}, F.~J. 2000, \mnras, 311, 433

\bibitem[{{Page} {et~al.}(1996){Page}, {Carrera}, {Hasinger}, {Mason},
  {McMahon}, {Mittaz}, {Barcons}, {Carballo}, {Gonzalez-Serrano}, \&
  {Perez-Fournon}}]{page1996}
{Page}, M.~J., {Carrera}, F.~J., {Hasinger}, G., {et~al.} 1996, \mnras, 281,
  579

\bibitem[{{Ranalli} {et~al.}(2013){Ranalli}, {Comastri}, {Vignali}, {Carrera},
  {Cappelluti}, {Gilli}, {Puccetti}, {Brandt}, {Brunner}, {Brusa},
  {Georgantopoulos}, {Iwasawa}, \& {Mainieri}}]{cdfscat}
{Ranalli}, P., {Comastri}, A., {Vignali}, C., {et~al.} 2013, \aap, 555, A42

\bibitem[{{Ranalli} {et~al.}(2015){Ranalli}, {Georgantopoulos}, {Corral},
  {Koutoulidis}, {Rovilos}, {Carrera}, {Akylas}, {Del Moro}, {Georgakakis},
  {Gilli}, \& {Vignali}}]{xmmatlas}
{Ranalli}, P., {Georgantopoulos}, I., {Corral}, A., {et~al.} 2015, \aap, 577,
  A121

\bibitem[{{Salvato} {et~al.}(2009){Salvato}, {Hasinger}, {Ilbert}, {Zamorani},
  {Brusa}, {Scoville}, {Rau}, {Capak}, {Arnouts}, {Aussel}, {Bolzonella},
  {Buongiorno}, {Cappelluti}, {Caputi}, {Civano}, {Cook}, {Elvis}, {Gilli},
  {Jahnke}, {Kartaltepe}, {Impey}, {Lamareille}, {Le Floc'h}, {Lilly},
  {Mainieri}, {McCarthy}, {McCracken}, {Mignoli}, {Mobasher}, {Murayama},
  {Sasaki}, {Sanders}, {Schiminovich}, {Shioya}, {Shopbell}, {Silverman},
  {Smol{\v c}i{\'c}}, {Surace}, {Taniguchi}, {Thompson}, {Trump}, {Urry}, \&
  {Zamojski}}]{salvato2009}
{Salvato}, M., {Hasinger}, G., {Ilbert}, O., {et~al.} 2009, \apj, 690, 1250

\bibitem[{{Salvato} {et~al.}(2011){Salvato}, {Ilbert}, {Hasinger}, {Rau},
  {Civano}, {Zamorani}, {Brusa}, {Elvis}, {Vignali}, {Aussel}, {Comastri},
  {Fiore}, {Le Floc'h}, {Mainieri}, {Bardelli}, {Bolzonella}, {Bongiorno},
  {Capak}, {Caputi}, {Cappelluti}, {Carollo}, {Contini}, {Garilli}, {Iovino},
  {Fotopoulou}, {Fruscione}, {Gilli}, {Halliday}, {Kneib}, {Kakazu},
  {Kartaltepe}, {Koekemoer}, {Kovac}, {Ideue}, {Ikeda}, {Impey}, {Le Fevre},
  {Lamareille}, {Lanzuisi}, {Le Borgne}, {Le Brun}, {Lilly}, {Maier},
  {Manohar}, {Masters}, {McCracken}, {Messias}, {Mignoli}, {Mobasher}, {Nagao},
  {Pello}, {Puccetti}, {Perez-Montero}, {Renzini}, {Sargent}, {Sanders},
  {Scodeggio}, {Scoville}, {Shopbell}, {Silvermann}, {Taniguchi}, {Tasca},
  {Tresse}, {Trump}, \& {Zucca}}]{salvato2011}
{Salvato}, M., {Ilbert}, O., {Hasinger}, G., {et~al.} 2011, \apj, 742, 61

\bibitem[{{Santini} {et~al.}(2009){Santini}, {Fontana}, {Grazian}, {Salimbeni},
  {Fiore}, {Fontanot}, {Boutsia}, {Castellano}, {Cristiani}, {de Santis},
  {Gallozzi}, {Giallongo}, {Menci}, {Nonino}, {Paris}, {Pentericci}, \&
  {Vanzella}}]{santini2009}
{Santini}, P., {Fontana}, A., {Grazian}, A., {et~al.} 2009, \aap, 504, 751

\bibitem[{{Schmidt}(1968)}]{schmidt68}
{Schmidt}, M. 1968, \apj, 151, 393

\bibitem[{{Schmidt} \& {Green}(1983)}]{schmidt1983}
{Schmidt}, M. \& {Green}, R.~F. 1983, \apj, 269, 352

\bibitem[{{Setti} \& {Woltjer}(1989)}]{sw89}
{Setti}, G. \& {Woltjer}, L. 1989, \aap, 224, L21

\bibitem[{{Silverman} {et~al.}(2008){Silverman}, {Green}, {Barkhouse}, {Kim},
  {Kim}, {Wilkes}, {Cameron}, {Hasinger}, {Jannuzi}, {Smith}, {Smith}, \&
  {Tananbaum}}]{silverman2008}
{Silverman}, J.~D., {Green}, P.~J., {Barkhouse}, W.~A., {et~al.} 2008, \apj,
  679, 118

\bibitem[{Skilling(2004)}]{skilling2004}
Skilling, J. 2004, Bayesian inference and maximum entropy methods in science
  and engineering, 735, 395

\bibitem[{Skilling {et~al.}(2006)}]{skilling2006}
Skilling, J. {et~al.} 2006, Bayesian Analysis, 1, 833

\bibitem[{{Treister} \& {Urry}(2006)}]{treister2006}
{Treister}, E. \& {Urry}, C.~M. 2006, \apjl, 652, L79

\bibitem[{{Treister} {et~al.}(2009){Treister}, {Urry}, \&
  {Virani}}]{treister2009}
{Treister}, E., {Urry}, C.~M., \& {Virani}, S. 2009, \apj, 696, 110

\bibitem[{{Trotta}(2008)}]{trotta2008}
{Trotta}, R. 2008, Contemporary Physics, 49, 71

\bibitem[{{Ueda} {et~al.}(2014){Ueda}, {Akiyama}, {Hasinger}, {Miyaji}, \&
  {Watson}}]{ueda2014}
{Ueda}, Y., {Akiyama}, M., {Hasinger}, G., {Miyaji}, T., \& {Watson}, M.~G.
  2014, \apj, 786, 104

\bibitem[{{Ueda} {et~al.}(2003){Ueda}, {Akiyama}, {Ohta}, \& {Miyaji}}]{ueda03}
{Ueda}, Y., {Akiyama}, M., {Ohta}, K., \& {Miyaji}, T. 2003, \apj, 598, 886

\bibitem[{{Ueda} {et~al.}(2008){Ueda}, {Watson}, {Stewart}, {Akiyama},
  {Schwope}, {Lamer}, {Ebrero}, {Carrera}, {Sekiguchi}, {Yamada}, {Simpson},
  {Hasinger}, \& {Mateos}}]{ueda2008}
{Ueda}, Y., {Watson}, M.~G., {Stewart}, I.~M., {et~al.} 2008, \apjs, 179, 124

\bibitem[{{Vito} {et~al.}(2014){Vito}, {Gilli}, {Vignali}, {Comastri}, {Brusa},
  {Cappelluti}, \& {Iwasawa}}]{vito2014}
{Vito}, F., {Gilli}, R., {Vignali}, C., {et~al.} 2014, \mnras, 445, 3557

\bibitem[{Watanabe(2010)}]{watanabe2010}
Watanabe, S. 2010, The Journal of Machine Learning Research, 11, 3571

\bibitem[{{Yencho} {et~al.}(2009){Yencho}, {Barger}, {Trouille}, \&
  {Winter}}]{yencho2009}
{Yencho}, B., {Barger}, A.~J., {Trouille}, L., \& {Winter}, L.~M. 2009, \apj,
  698, 380

\bibitem[{{Zubovas} \& {King}(2012)}]{zubovas2012}
{Zubovas}, K. \& {King}, A.~R. 2012, \mnras, 426, 2751

\end{thebibliography}

\end{document}